\documentclass{article}

\usepackage[margin=2.5cm]{geometry}
\usepackage{graphicx,xcolor}
\DeclareGraphicsExtensions{.png,.jpg}
\usepackage{amsmath,amsfonts,amssymb,amsthm}
\usepackage{enumitem} 
\usepackage{url}
\usepackage{multirow}

\usepackage{hyperref}
\hypersetup{pdfstartview=FitH}
\hypersetup{bookmarks=false,bookmarksopen=false}  % doesn't work
\hypersetup{draft} % this works

\usepackage{tikz}
\usetikzlibrary{spline,arrows.meta,decorations.pathmorphing}

\usetikzlibrary{external}
\tikzset{
    png export/.style={
        external/system call/.add={}{; convert -density 300 -transparent white "\image.pdf" "\image.png"},
        /pgf/images/external info,
        /pgf/images/include external/.code={
          \includegraphics[width=\pgfexternalwidth,height=\pgfexternalheight]{##1.png}
        },
    }
}

\theoremstyle{plain}
\newtheorem{exercise}{Exercise}

\newcommand{\ds}{\displaystyle}
\newcommand{\azimuth}{\varphi}
\newcommand{\inclination}{\theta}
\newcommand{\latitude}{\lambda}
\newcommand{\longitude}{\psi}
\newcommand{\subtext}[2]{{#1}_{\text{\emph{#2}}}}
\newcommand{\Pmax}{\subtext{P}{max}}
\newcommand{\vwind}{\subtext{v}{wind}}
\newcommand{\acmin}{\subtext{a}{c,min}}
\newcommand{\gamskid}{\subtext{\gamma}{skid}}
\newcommand{\gamstep}{\subtext{\gamma}{step}}

\definecolor{mygreen}{rgb}{0,0.5,0}  % for skier diagram

\usepackage[normalem]{ulem}
\definecolor{myBrickRed}{rgb}{0.80,0.20,0.20}
\definecolor{myGray}{rgb}{0.7,0.7,0.7}

\newcommand{\Matlab}{MATLAB}
\begin{document}
\renewcommand{\thefootnote}{\fnsymbol{footnote}}
\title{A mathematical model for Nordic skiing%
  \footnote{%
    % Version of \today (arXiv).
    This work was supported by a Postdoctoral
    Fellowship (JSM) from the Pacific Institute for the Mathematical
    Sciences, and a Discovery Grant (JMS) from the Natural Sciences and
    Engineering Research Council of Canada.}}
\author{Jane Shaw MacDonald\footnote{Department of Mathematics,
    Oregon State University, Corvallis, OR, USA 
    ({\tt macdojan@oregonstate.edu}).} 
  \and Rafael Ordo\~nez Cardales\footnote{Universidad Popular del
    C\'esar, Departamento de Matem\'aticas y Estad\'isticas,
    Valledupar, C\'esar, Colombia ({\tt reordonezc@unicesar.edu.co}).}
  \and John M. Stockie\footnote{Department of Mathematics, 
    Simon Fraser University, Burnaby, BC, Canada ({\tt
      jstockie@sfu.ca}).}}  

\maketitle

\begin{abstract} 
  Nordic skiing provides fascinating opportunities for mathematical
  modelling studies that exploit methods and insights from physics,
  applied mathematics, data analysis, scientific computing and sports
  science. A typical ski course winds over varied terrain with frequent
  changes in elevation and direction, and so its geometry is naturally
  described by a three-dimensional space curve. The skier travels along
  a course under the influence of various forces, and their dynamics can
  be described using a nonlinear system of ordinary differential
  equations (ODEs) that are derived from Newton's laws of motion.  We
  develop an algorithm for solving the governing equations that combines
  Hermite spline interpolation, numerical quadrature and a high-order
  ODE solver.  Numerical simulations are compared with measurements of
  skiers on actual courses to demonstrate the effectiveness of the
  model.
  Throughout, we aim to illustrate how elementary concepts from
  undergraduate courses in calculus and scientific computing can be
  applied to study real problems in sport, which we hope will provide
  stimulating examples for both instructors and students.  At the same
  time, we demonstrate how these concepts are capable of providing novel
  insights into skiing that should also be of interest to sport
  scientists.
\end{abstract}



\renewcommand{\thefootnote}{\arabic{footnote}}

%%%%%%%%%%%%%%%%%%%%%%%%%%%%%%%%%%%%%%%%%%%%%%%%%%%%%%%%%%%%%%%%%%%%%% 
% --------------------------------------------------------------------
%%%%%%%%%%%%%%%%%%%%%%%%%%%%%%%%%%%%%%%%%%%%%%%%%%%%%%%%%%%%%%%%%%%%%% 

\section{Background}
\label{sec:intro}

Nordic skiing, also known as cross-country skiing, is a winter sport
that attracts participants at all fitness levels, ranging from
recreational skiers to athletes competing in World Cup and Olympic
races. Nordic ski courses or trails are groomed, snow-covered paths that
are carefully designed to provide a variety of terrain interspersing
flats with undulating stretches, moderate climbs with steep
up/down-hills, and gently winding turns with tight curves -- all of
which provide both casual skiers and serious athletes with something to
enjoy and challenge themselves. Compared to alpine skiing, where the
entire course is downhill and gravity does the work of propulsion,
Nordic skiers spend a majority of their time on flats and uphills where
they must maintain their forward motion by generating propulsive forces
using a combination of techniques that engage their entire body,
including arms, legs, and torso. As a result, Nordic skiing is
frequently touted as an ideal activity to foster aerobic
fitness~\cite{hutchinson-2018, moxnes-hausken-2009} and even reducing
mortality~\cite{laukkanen-etal-2018}, with St\"oggl
al.~\cite{stoggl-etal-2016} concluding that ``cross-country skiing can
be regarded as the gold standard winter time aerobic exercise mode, with
a high percentage of muscles in the whole body being activated and the
highest VO2max values among all sports.''
% From Moxnes and Hausken (2009): ``Skis are far superior to using
% snowshoes or walking.  Skiing is also a valuable recreational
% sport. The skier spends his day in open air, usually in beautiful
% scenery far away from the dust and the noisy atmosphere of cities. All
% age groups can participate if the track and velocity are chosen
% according to skill and age.  The feeling of exertion at a given oxygen
% consumption is less than for many other types of exercises,
% e.g. walking, running or bicycling. The reason is that the effort
% during skiing is spread to a large number of powerful muscles
% throughout the body, not only the leg muscles, and they work under
% relatively favourable conditions, due to alteration of work and
% relative rest during glide. The consequence of this is that an athlete
% who wants to improve his general condition, especially his maximum
% capacity of respiration and circulation, can use skiing as a
% conditioning exercise, even if he wants to compete in discipline other
% than skiing.''

Beyond its athletic appeal, Nordic skiing also provides diverse
opportunities for mathematical modelling studies that combine aspects
from course geometry, biomechanics, skier dynamics (governed by muscle
propulsion and snow/air resistance), and race strategy. The complex
terrain and highly variable snow conditions on a typical ski course
induce skiers to employ a variety of techniques (double poling, diagonal
striding, step or skid turns, etc.) that add further
complexity~\cite{moxnes2008cross}.
% Among many nice quotes: ``Skiing performance on flat terrain is very
% different from walking and running performance on flat terrain.''
This is in contrast with other sports like running, cycling and
swimming, where the technique used varies less within a race and is
determined almost solely by race distance rather than variations in the
course~\cite{keller1974optimal, maronski-1996}. In this paper, we
present a model grounded in Newtonian physics that captures the balance
of forces acting on an athlete skiing in both two and three dimensions,
which is formulated as a system of nonlinear ordinary differential
equations (ODEs).  This problem is suitable for treatment within a
variety of undergraduate classes in mathematical modelling, ordinary
differential equations, or scientific computing.

Sport scientists have already devoted significant effort to
investigating the dynamics of Nordic skiing, leading to many studies in
which differential equation models have been used to study questions
related to athlete propulsion and optimizing race strategies. Our work
is largely based on the model of Moxnes, Sandbakk and
Hausken~\cite{moxnes2014using} (which we abbreviate as MSH), who derived
a system of two ODEs that capture the motion of an athlete along a 2D
race course in response to forces of propulsion, gravity, snow
resistance and aerodynamic drag. Similar ODE-based models of skier
dynamics have also appeared in~\cite{carlsson-chapter-2016,
  carlsson-etal-2011, moxnes2008cross, ni-liu-zhang-ke-2022,
  sundstrom2013numerical}, whereas other researchers have developed much
more complicated biomechanical models that capture the detailed arm, leg
and body motions employed in specific
techniques~\cite{andersson-etal-2014, Driessel2004dynamics,
  moxnes-hausken-2009}.

Our main goal is to provide a detailed derivation of the coupled ODE
system for the 2D MSH model, along with an associated numerical
scheme. We propose several modifications and improvements to this model
that are easily accessible to students at the undergraduate level:
\begin{itemize}
\item We replace the piecewise linear course geometry from
  MSH~\cite{moxnes2014using} with a smoother cubic
  spline interpolant that more accurately captures the shape of a real
  ski course (see Fig.~\ref{fig:spline} for an example).
\item We constrain the skier to move along the interpolated path, in
  contrast with some models that introduce numerical errors appearing as
  a ``drift'' away from the actual course.
\item The governing equations are solved using a state-of-the art
  numerical ODE solver (\Matlab's {\tt ode113}) that incorporates
  high-order derivative approximations, adaptive time stepping, error
  control and event detection.
\item The arc length integral for skier distance is approximated using
  Simpson's rule which is a fourth-order numerical quadrature formula.
\item We incorporate a realistic 3D course (such as in
    Fig.~\ref{fig:FISOle}a) that includes track curvature, which allows
  us to simulate different braking techniques used by athletes to safely
  negotiate tightly-curved downhill
  sections~\cite{sandbakk-supej-etal-2013}.  This is important in race
  settings where athletes strive for any possible advantage over their
  competitors.
\end{itemize}
Taking advantage of \Matlab's built-in interpolation algorithms and ODE
solvers~\cite{matlab-r2023a}, we provide numerical simulations that
clearly demonstrate the advantages of these model extensions. Taken
together, this study demonstrates many natural connections with standard
material taught in the undergraduate math curriculum, in courses such as
vector calculus, ordinary differential equations, numerical methods, and
mathematical modelling. Consequently, we hope to convince the reader
that relatively elementary mathematical concepts and techniques can give
rise to novel results that have practical application to real problems
in sport.

%%%%%%%%%%%%%%%%%%%%%%%%%%%%%%%%%%%%%%%%%%%%%%%%%%%%%%%%%%%%%%%%%%%%%%
\subsection{Exercises, \Matlab\ Codes and Data Files}

Exercises are distributed throughout this paper to provide opportunities
to study missing details, to dig more deeply into examples, and to
further explore the material.  Solutions to all exercises are provided
as Supplementary Material which can be downloaded from this
website:
\begin{center}
  \url{http://www.sfu.ca/~jstockie/skiing/}
\end{center}
All data files and \Matlab\ codes used to generate the numerical results
are also provided with the Supplementary Materials.  Note that these
codes make use of two functions that are not part of the standard
\Matlab\ distribution: {\tt fnder} in the Curve Fitting
Toolbox~\cite{matlab-curvefit-2023} and {\tt gpxread} in the Mapping
Toolbox~\cite{matlab-mapping-2023}.

%%%%%%%%%%%%%%%%%%%%%%%%%%%%%%%%%%%%%%%%%%%%%%%%%%%%%%%%%%%%%%%%%%%%%%
% --------------------------------------------------------------------
%%%%%%%%%%%%%%%%%%%%%%%%%%%%%%%%%%%%%%%%%%%%%%%%%%%%%%%%%%%%%%%%%%%%%% 

\section{Constructing a Smooth Ski Course from Sparse Data}
\label{sec:course-spline}

Generating realistic simulations of skier dynamics requires a similarly
realistic representation of the path that they ski along.  However, real
ski course geometries are typically only available in the form of a 2D
elevation plot (of height versus distance) or in some cases as a GPS
file that defines a relatively sparse sample of waypoints along the
course.  As a result, before we even start to develop a model, it is
important to first grapple with several issues related to representing
the course itself:
\begin{itemize}
\item Selecting an appropriate parameterization so that the ski course
  can be represented (in 2D or 3D) as a continuous real-valued function
  for use in an ODE model.
\item Generating a suitable smooth approximation for a sparse set of
  data points that remains as faithful as possible to the actual course
  underlying the given data.
\item Verifying that the final parameterized curve satisfies the
  specifications for an official race course, and dealing with any gaps
  or errors in the data.
\end{itemize}

%%%%%%%%%%%%%%%%%%%%%%%%%%%%%%%%%%%%%%%%%%%%%%%%%%%%%%%%%%%%%%%%%%%%%%
\subsection{Parameterizing a Ski Course}
\label{sec:space-curves}

\begin{figure}[tb]
  \centering
  \begin{tabular}{cc}
     (a) Ole course (3D view) & (b) 2D elevation profile\\
     \begin{tikzpicture}
       \node at (0,0) {\includegraphics[width=0.45\textwidth,trim=0 0 0 40,clip]{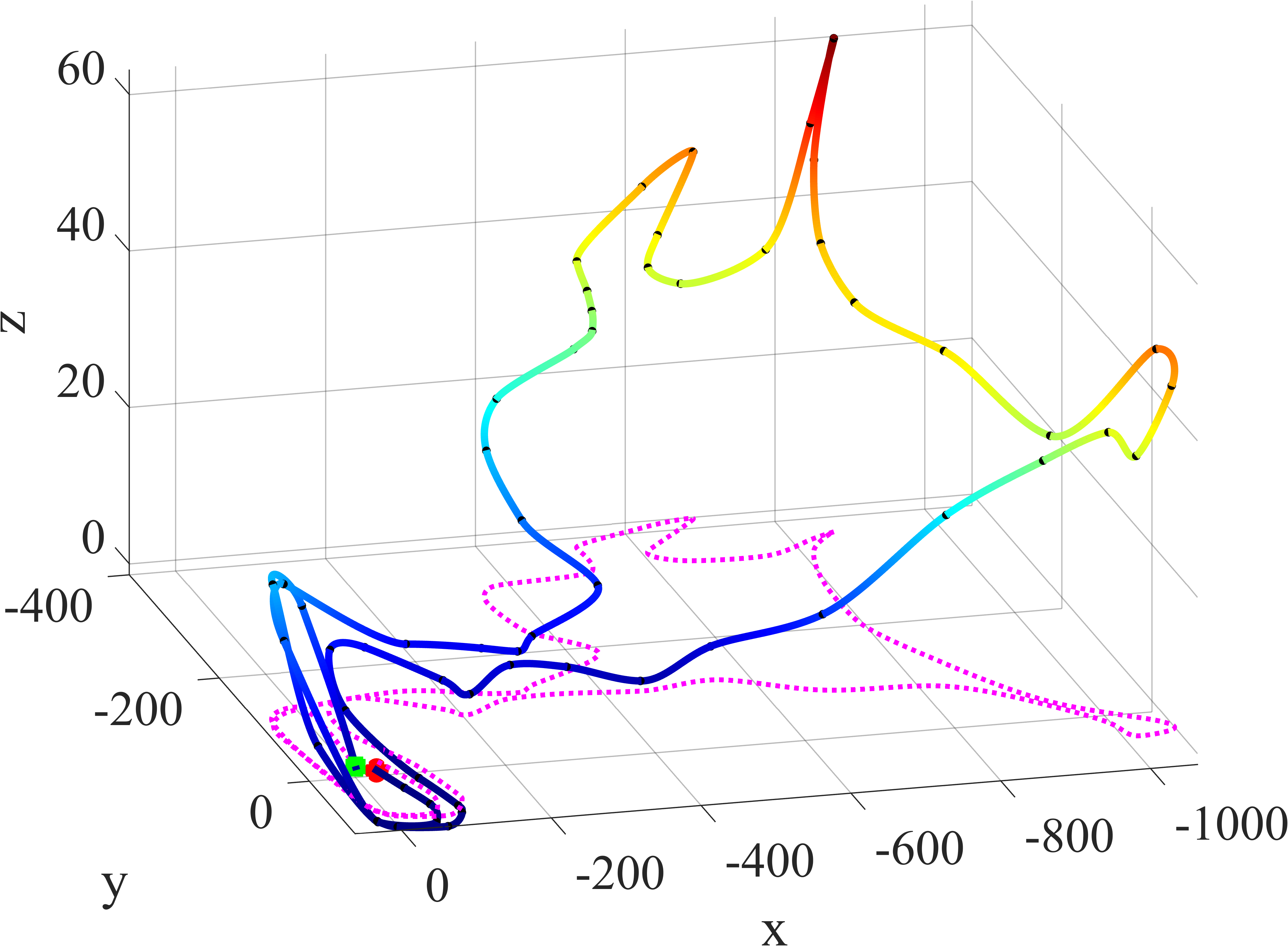}};
       \draw[black!50!white,thick,dash pattern=on 2pt off
       1pt,rotate=-28] (1.9,1.85) ellipse (0.8 and 0.6); 
       \draw[black!50!white,thick,dash pattern=on 2pt off
       1pt,rotate=-15] (2.6,-0.35) ellipse (0.8 and 0.3); 
     \end{tikzpicture}
     & 
     \begin{tikzpicture}
       \node at (0,0) {\includegraphics[width=0.45\textwidth]{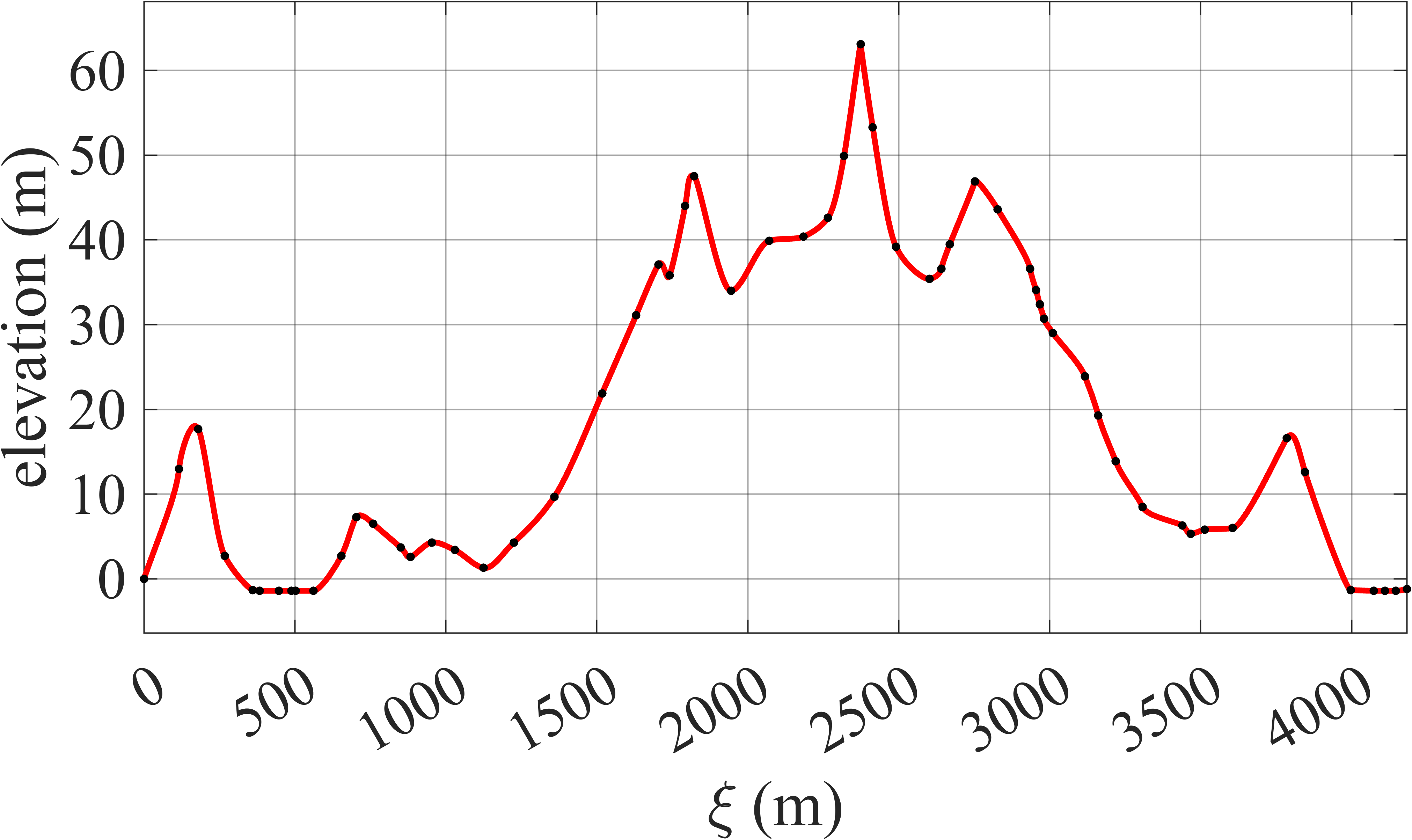}};
       \draw[black!50!white,thick,dash pattern=on 2pt off 1pt] (0.1,1)
       ellipse (0.5 and 0.5); 
        %\draw[black!50!white,thick,dash pattern=on 2pt off 1pt]
        %(-0.3,0.6) -- (-0.3,1.5) -- (0.55,1.5) -- (0.55,0.6) --
        %(-0.3,0.6); 
     \end{tikzpicture}
  \end{tabular}\\
  (c) Three parameterizations for elevation\\
  \includegraphics[width=0.5\textwidth]{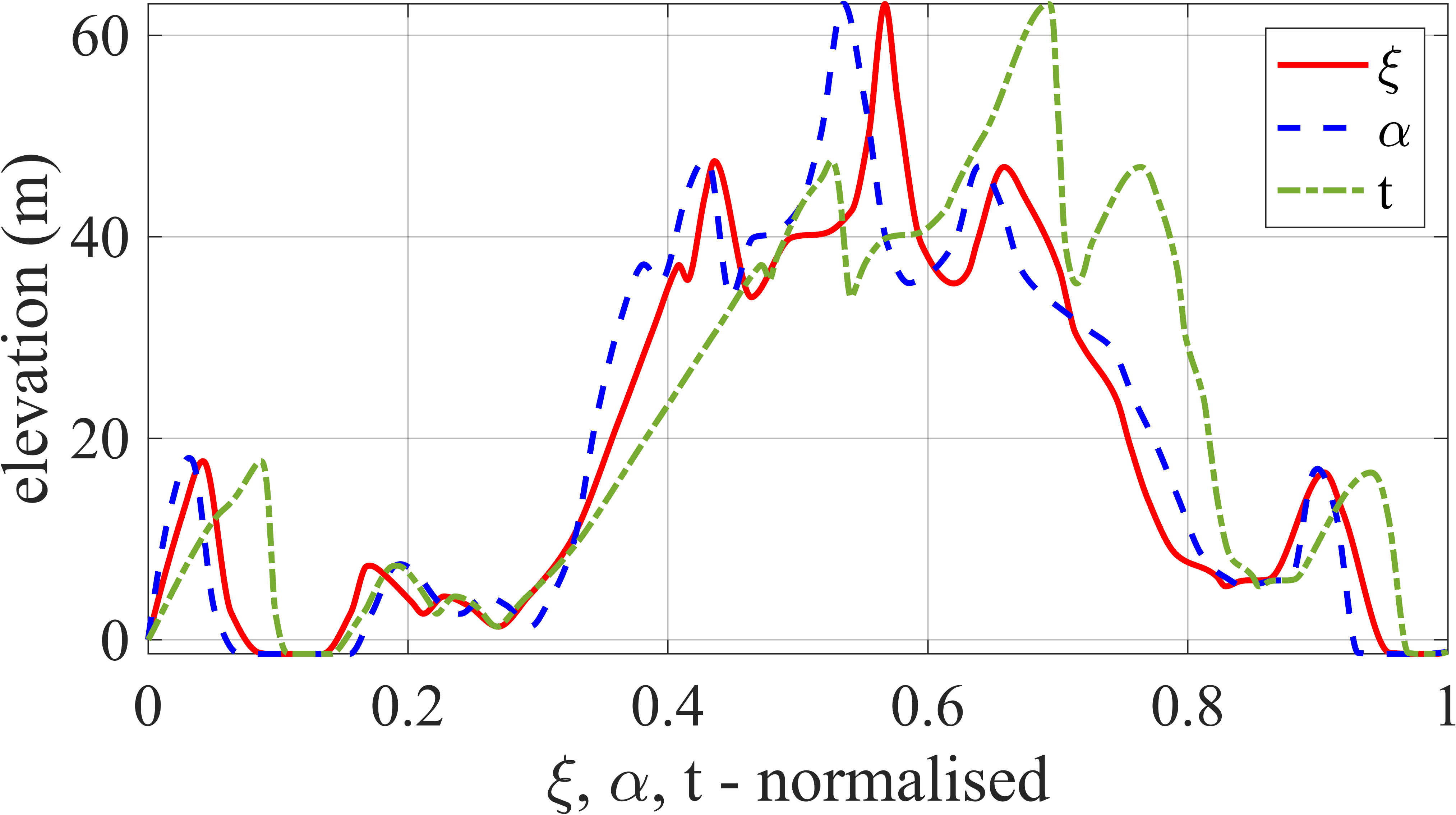}
  \caption{Three views of the 4.2~km Ole course in
    Toblach~\cite{dolomiti-gps-2023}. (a,~Top~Left) 3D plot of the skier
    path defined by 58 GPS points $\vec{r}_i$ for $i=0,1,2,\dots, 57$
    (black points), interpolated by a smooth Hermite spline curve.  The
    start and finish in the left foreground are indicated by a green
    square and purple circle. The dotted magenta curve is the projection
    of the course in the horizontal plane. (b,~Top~Right) An elevation
    profile plot, which is an ``unwound'' view of the course in 2D and
    depicts elevation $z$ as a function of horizontal distance $\xi$,
    where $\xi$ is arc length along the projected skier path (i.e., the
    magenta curve from~(a)).  The dashed grey ovals refer to the zoomed
    plots in Fig.~\ref{fig:spline}. (c,~Bottom) Comparison of three
    parameterizations for elevation: $\xi$ (red, solid), $\alpha$ (blue,
    dashed) and $t$ (green, dash-dot), with each parameter normalized to
    lie in $[0,1]$.}
  \label{fig:FISOle}
\end{figure}
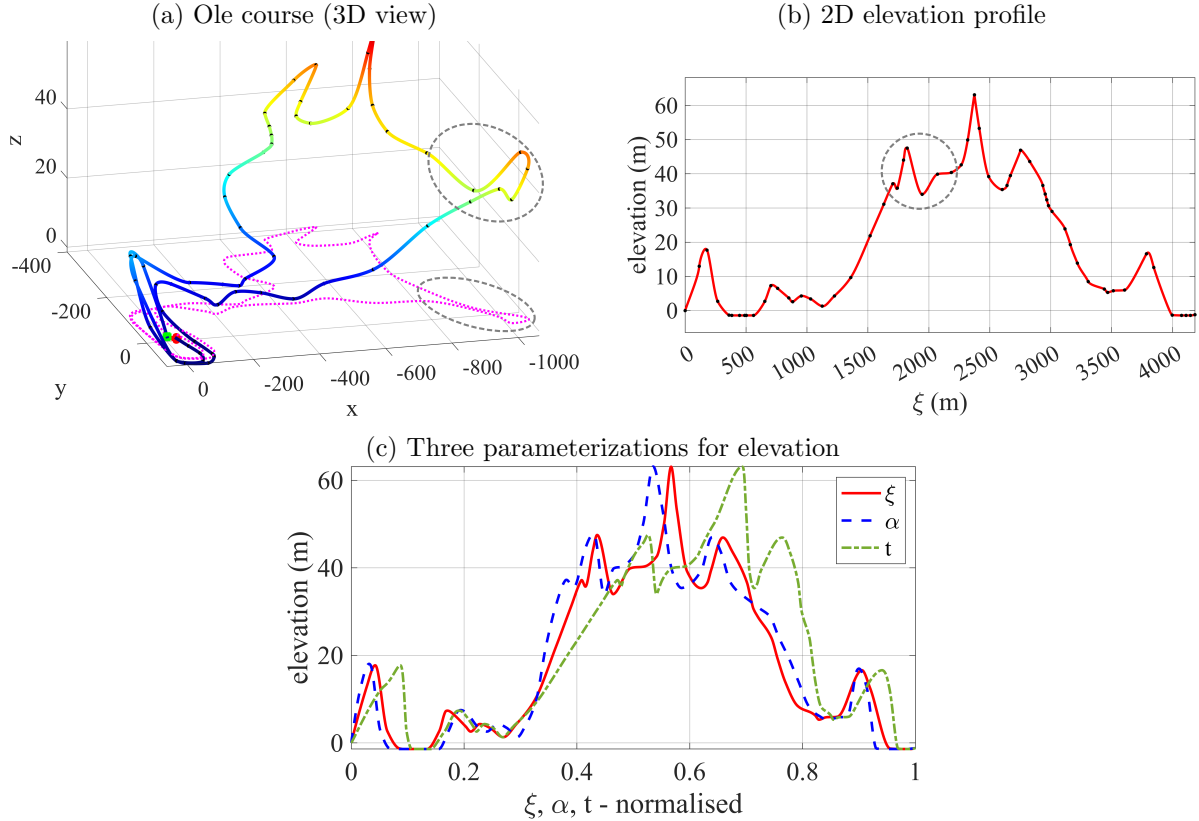

The path that a skier follows while propelling themselves along a real
ski course is a 3D curve such as that depicted in
Fig.~\ref{fig:FISOle}a. This particular plot is generated from GPS
data~\cite{dolomiti-gps-2023} measured on the 4.2~km ``Ole'' course in
Toblach, Italy. This course has been officially certified for
competition purposes by FIS (F\'ed\'eration Internationale de Ski) and
has been the location of many high-level races including the 2024 FIS
World Cup Tour de Ski.
%% https://www.dolomitinordicski.com/en/regions/3-zinnen-dolomites/events/1-coop-fis-cross-country-world-cup.html
The raw GPS data consists of 58 measurements of latitude $\latitude$,
longitude $\longitude$, and elevation $z$, which are read using the
\Matlab\ function {\tt gpxread} and must first be converted to Cartesian
coordinates on a flat patch of the Earth's surface. The curve in
Fig.~\ref{fig:FISOle}a is plotted in terms of $(x,y,z)$, where $z$ is
elevation, and $x,y$ lie in the horizontal plane%
\footnote{When latitude $\latitude$ and longitude $\longitude$ are
  measured in degrees, the north-south distance (in m) is
  $y\approx 111133\longitude$ whereas the east-west distance is
  $x\approx 111413 \latitude \cos\left(\pi\latitude/180\right)$. The
  latter contains an extra cosine adjustment factor because the distance
  between lines of longitude varies with $\latitude$. These two formulas
  are actually the leading order terms in an approximation that accounts
  for the oblate spheroidal shape of the Earth, as explained in the
  \href{https://en.wikipedia.org/wiki/Geographic_coordinate_system}{Wikipedia
    article ``Geographic coordinate system.''}}.
% Exact formulas from Wikipedia (distance correct to within 1 cm):  
% yfac = 111132.92 - 559.82 \cos(2\phi) + 1.175 \cos(4\phi) - 0.0023 \cos(6\phi)
% xfac = 111412.84 \cos(\phi) - 93.5 \cos(3\phi) + 0.118 \cos(5\phi)
% 
For convenience, we also translate coordinates so that the course starts
at the origin (highlighted in the plot by a green square).  A second
view of the course is obtained in Fig.~\ref{fig:FISOle}b by
``unwinding'' or straightening this 3D curve and displaying it as a 2D
plot of elevation versus a new distance variable $\xi$, which represents
the arc length measured along the projected path in the $x,y$-plane (the
dotted magenta curve in Fig.~\ref{fig:FISOle}a). This 2D elevation plot
of $(\xi,z)$ is the official method for reporting course elevation
profiles in competition documents (private communication, Al
  Maddox, Canadian FIS Technical Delegate, June 9, 2024).

This example highlights the importance of exploiting different
parameterizations to represent a skier's path. In general, any space
curve may be expressed in vector form as
$\vec{r}(\alpha) = \big(x(\alpha),\, y(\alpha),\, z(\alpha)\big)$, where
$\alpha$ is a parameter that in this paper will be chosen in three ways:
\begin{description}
\item[{\normalfont\itshape -- Time parameterization:}] When modelling
  the dynamics of a skier whose path varies in time $t$, we view the
  position $\vec{r}(t)=(x(t), y(t), z(t))$ as being parameterized by
  time.

\item[{\normalfont\itshape -- Distance parameterization:}] Ski course
  geometry is most commonly available as a 2D plot of elevation versus
  horizontal distance such as in Fig.~\ref{fig:FISOle}b, where elevation
  is viewed as a function $z(\xi)$ parameterized by ``projected arc
  length'' $\xi$ (the distance along the skier path projected in the
  horizontal plane).

\item[{\normalfont\itshape -- Arbitrary parameterization:}] When the
  course is specified using $N+1$ GPS points with coordinates
  $(x_i, y_i, z_i)$ for $i=0,1,\dots, N$, it is convenient to base the
  parameter on the index $i$. In this case, we represent the curve as
  $\vec{r}(\alpha) = (x(\alpha), y(\alpha), z(\alpha))$ where
  $\alpha\in[0,1]$ and GPS points correspond to discrete values of the
  parameter $\alpha_i = \frac{i}{N}$ (for simplicity, we use
  equally-spaced points on the unit interval).
\end{description}
In all three cases, we employ the same notation $\vec{r}=(x,y,z)$ to
denote position, and make use of different parameterizations as
needed. The effect of different parameter choices is nicely illustrated
in Fig.~\ref{fig:FISOle}c, which displays the projected arc length curve
$z(\xi)$ from Fig.~\ref{fig:FISOle}b along with $z(\alpha)$ and
$z(t)$. All three parameters have been re-scaled to lie in $[0,1]$ in
order to demonstrate how the ``physical'' elevation plot $z(\xi)$
becomes distorted when displayed using one of the other two
parameterizations.

%%%%%%%%%%%%%%%%%%%%%%%%%%%%%%%%%%%%%%%%%%%%%%%%%%%%%%%%%%%%%%%%%%%%%%
\subsection{Interpolating with Cubic Hermite Splines}
\label{sec:splines}

A dynamic model for a skier moving along a smooth course should
determine the location and speed of the skier at any time $t$.  However,
a real course geometry is only defined at a discrete set of points,
either obtained from a 2D elevation profile plot (like
Fig.~\ref{fig:FISOle}b) or a 3D data file in GPS Exchange Format.
Therefore, a fundamental first step in the modelling process is to
construct a smooth (2D or 3D) curve that passes through these data
points.  This is a standard problem in \emph{data
  interpolation}.

With a bit of foresight, it should be clear that any model of Nordic
skiing will require quantities such as slope (or inclination angle) and
curvature at the skier's position $\vec{r}(\alpha)$, parameterized by
$\alpha\in[0,1]$. Slope and curvature depend on the first two
derivatives of position, $\vec{r}_\alpha$ and $\vec{r}_{\alpha\alpha}$,
so that any interpolant we use must be twice differentiable. This
suggests using a \emph{spline interpolant}, which is a piecewise-defined
function that interpolates each successive pair of data points with a
polynomial (refer to the classic text by de Boor for more
  information~\cite{deboor-1978}).  For a 2D course a single
interpolant $z=Z(\alpha)$ will suffice, but in 3D we will need to
construct three spline interpolants, one for each of the coordinate
functions: $x=X(\alpha)$, $y=Y(\alpha)$ and $z=Z(\alpha)$.

\begin{figure}[tb]
  \centering
  \begin{tabular}{ccc}
    (a) Elevation view
    && (b) Plan (overhead) view \\
    \raisebox{1cm}{\includegraphics[width=0.47\textwidth]{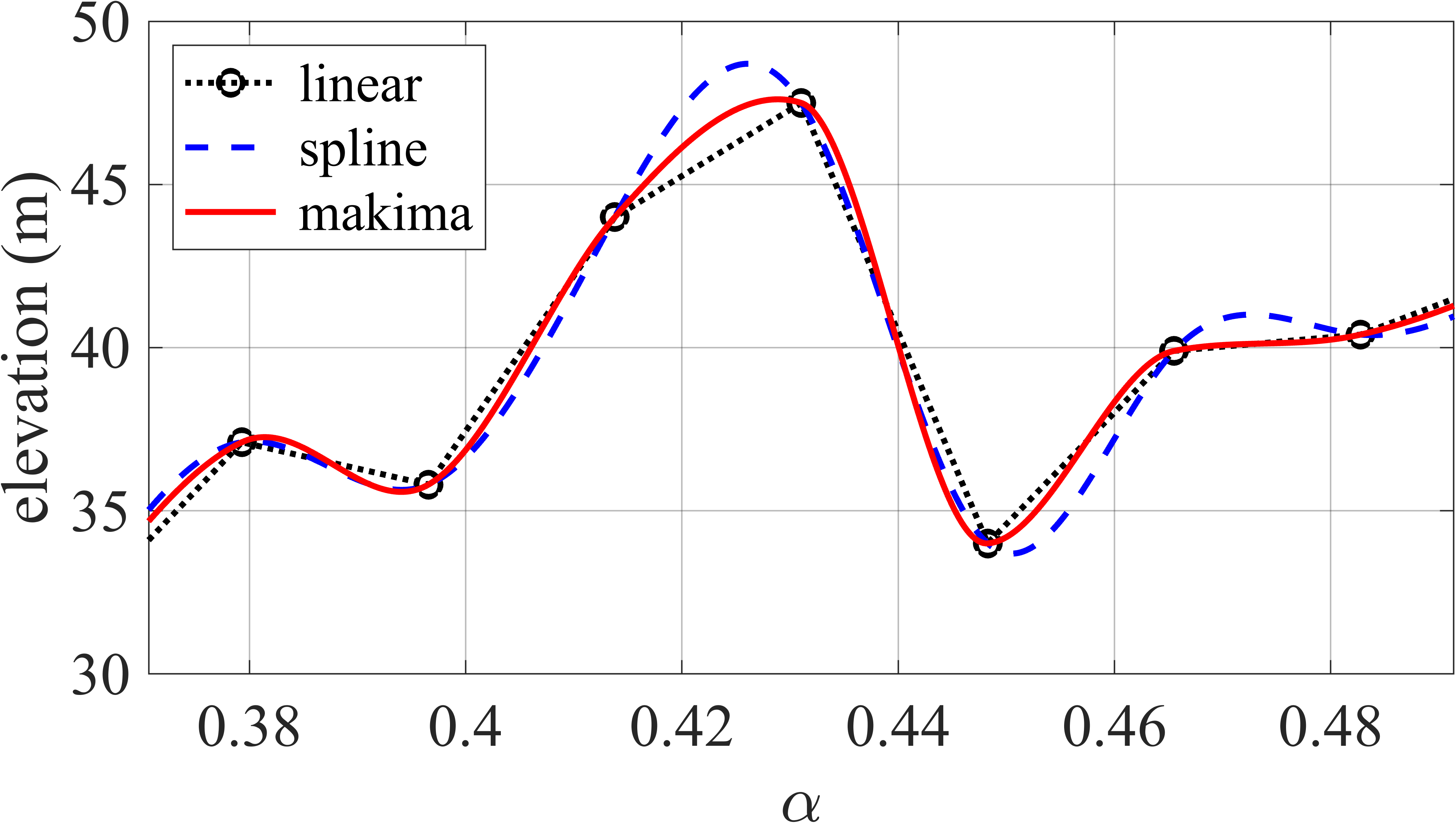}}
    && \includegraphics[width=0.41\textwidth]{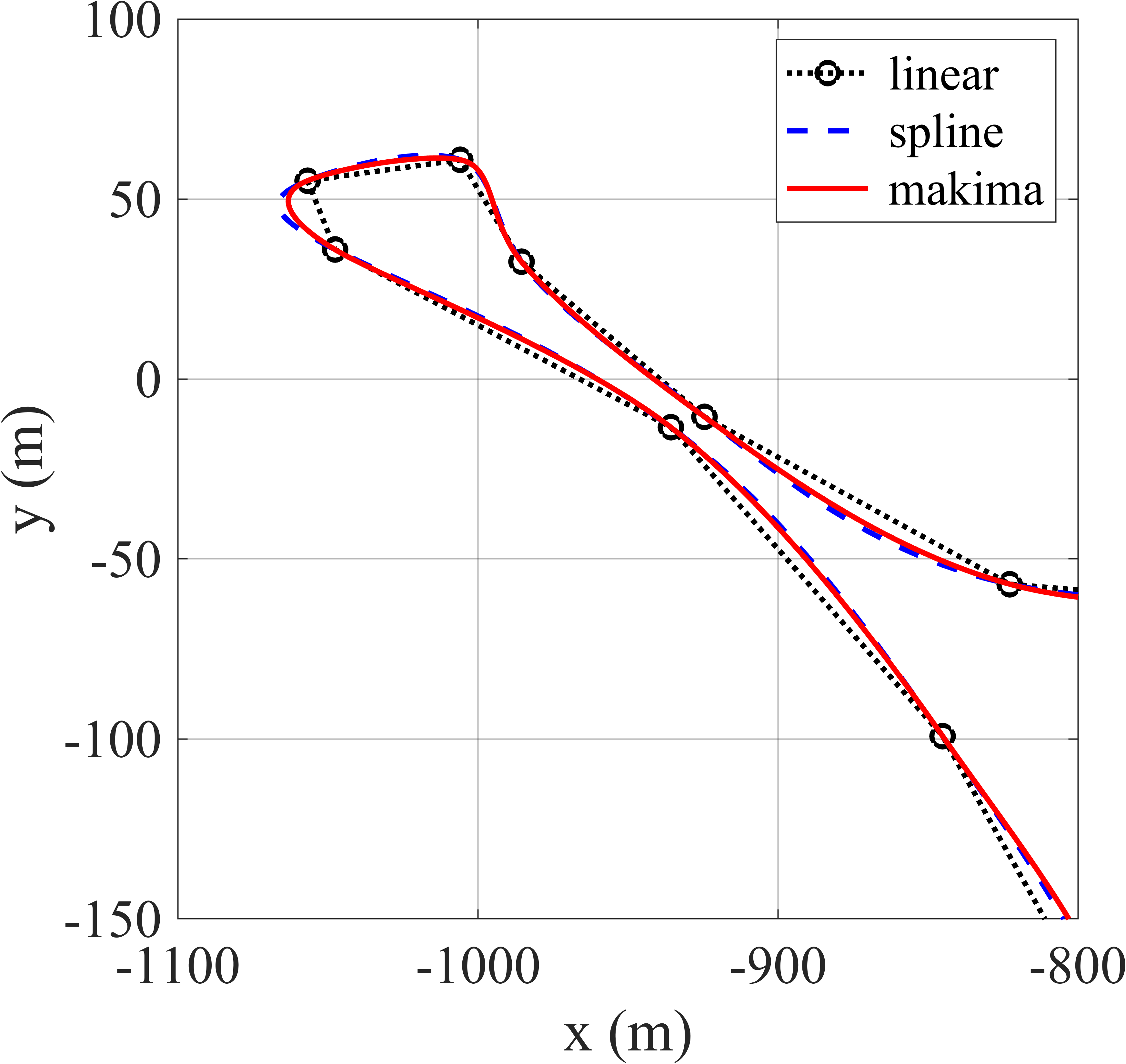}
  \end{tabular}
  \caption{Comparison of three splines using both elevation (a,left) and
    plane views (b,right): linear (dotted black), cubic (dashed blue)
    and cubic Hermite (solid red). The GPS data points are indicated
    (with black circles) and each spline is plotted using a fine grid of
    4000 equally-spaced $\alpha$-points. Both plots show the same
    ``zoomed-in'' section of the Ole ski course, for GPS data points
    $\vec{r}_i$ numbered $i=22, \dots, 28$ (this is the circled region
    in Fig.~\ref{fig:FISOle}).}
  \label{fig:spline}
\end{figure}

The simplest choice of interpolant is a linear spline that connects each
pair of points with a straight line segment, which is depicted in
Fig.~\ref{fig:spline} in the zoomed-in views of a short section of the
Ole course. This interpolant is continuous but only piecewise
differentiable, meaning that the derivative (slope) is only piecewise
continuous since it is undefined at the data points!  Furthermore,
curvature is identically zero along all linear segments, so it is
impossible to capture any curvature-dependent effects related to skier
turning dynamics.

\begin{figure}[tb]
  \centering
  \includegraphics[width=0.5\textwidth]{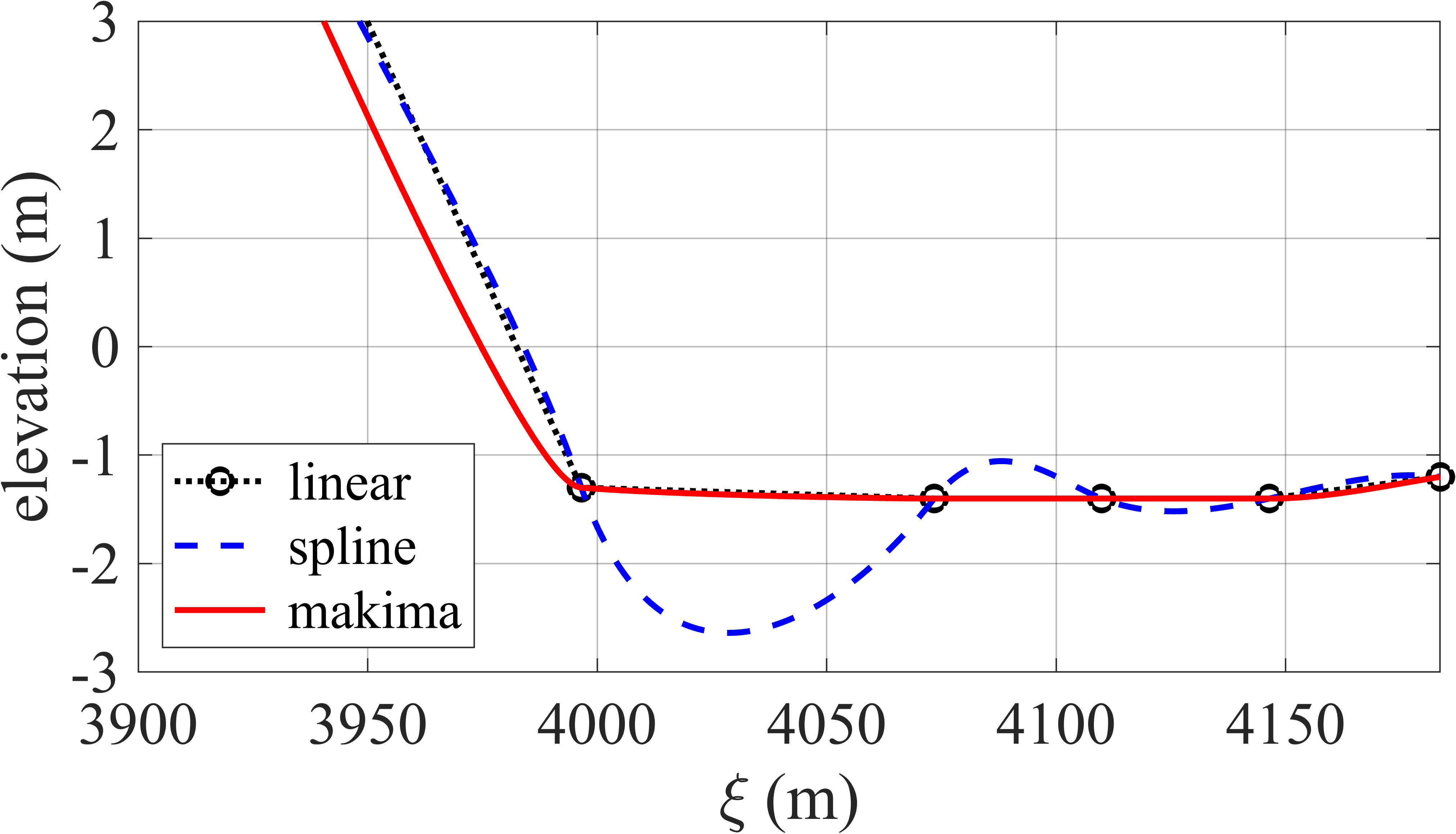}
  \caption{The final 250~m stretch of the Ole course, emphasizing the
    spurious oscillations that can be generated by a cubic spline when
    points are relatively flat and closely-spaced.}
  \label{fig:zoom-finish}
\end{figure}

Another smoother alternative is the usual cubic spline, which is
constructed so that the function and its first two derivatives are
continuous at all interpolation points. Fig.~\ref{fig:spline} displays a
zoomed-in view of the cubic spline interpolant for the Ole course data
(computed with \Matlab's {\tt spline} function).  The resulting curve is
clearly smoother than the linear spline, but reveals a disadvantage of
the cubic spline: namely, that fitting a smooth interpolant to
oscillatory data like a ski course can generate over- and under-shoots
that appear visually inconsistent with the original data.  These
spurious oscillations arising in polynomial interpolation are referred
to as Runge's phenomenon, or more informally as ``polynomial wiggle.''
This example highlights a problem that is further exacerbated by the
irregular spacing between GPS points, which varies from 13--179~m and is
easily seen in Fig~\ref{fig:FISOle}b. When fitting a cubic spline to
such data, the amplitude of these oscillations tend to be magnified
where interpolation points are closest. This is most evident at the
finish where the points are especially tightly clustered, leading to the
unrealistic oscillations shown in the close-up view in
Fig.~\ref{fig:zoom-finish}.

Ideally, we seek an interpolant that is smooth (at least differentiable)
and also avoids creating any new oscillations. This is precisely the aim
of Hermite cubic interpolation (implemented in \Matlab's {\tt makima}
function), which relaxes the requirement that the second derivative be
continuous and instead ``respects monotonicity and is
shape-preserving''~\cite{matlab-r2023a}\footnote{\Matlab\ has two
  functions that implement Hermite cubic splines: {\tt pchip} and {\tt
    makima}. A detailed comparison of these two interpolants with the
  usual cubic spline is provided in two \Matlab\ Blog posts by
  Moler~\cite{moler-pchip-2012, moler-makima-2019}.  We choose the
  ``modified Akima'' Hermite interpolant {\tt makima} because (according
  to Moler) ``it produces undulations which find a nice middle ground
  between `{\tt spline}' and `{\tt pchip}'{},'' which in our experience
  means that it also generates the best results when applied to ski
  course data.}. The Hermite spline interpolant pictured in
Figs.~\ref{fig:spline} and~\ref{fig:zoom-finish} clearly eliminates the
over/under-shoots in the cubic spline and more closely follows trends in
the data; however, this comes at the expense of losing one degree of
smoothness at the data points.

Another side-effect of the highly variable spacing between GPS points is
that the equal-$\alpha$ parameterization in Fig.~\ref{fig:spline}a
distorts the elevation curve, as shown in
Fig.~\ref{fig:FISOle}c. Consequently, the projected arc length
parameterization $z(\xi)$ in Fig.~\ref{fig:FISOle}b is a more
``natural'' choice for a ski course because it provides an accurate
visual representation of geometric quantities like distance, slope and
curvature.

In practice, we do not observe any evidence of significant noise or
random variation in the course data, and except for certain isolated
anomalies discussed later in Section~\ref{sec:homologation}, the given
point locations seem to be generally reliable. This suggests that
interpolating splines are a better choice than alternative curve fitting
options such as B-splines or other smoothing splines. Nonetheless, we
encourage the interested reader to explore the use of other functions
like {\tt bspline} and {\tt csaps} in \Matlab's Curve Fitting Toolbox.

%%%%%%%%%%%%%%%%%%%%%%%%%%%%%%%%%%%%%%%%%%%%%%%%%%%%%%%%%%%%%%%%%%%%%%
\subsection{Computing Distance and Inclination Angle}
\label{sec:distances}

We are now prepared to address the distinction between modelling racers
skiing on 2D and 3D courses: in 2D, where the geometry is given by an
elevation profile plot of $z$ versus $\xi$; and in 3D, where $(x,y,z)$
coordinates are obtained from GPS data. In the 2D case, points
$(\xi_i,z_i)$ can be obtained from an elevation profile plot with a data
extraction tool such as WebPlotDigitizer~\cite{webPlotDigitizer}, and
then used to construct the parametric splines $\Xi(\alpha)$,
$Z(\alpha)$. In 3D, the GPS points $(x_i, y_i, z_i)$ are interpolated to
obtain the splines $X(\alpha)$, $Y(\alpha)$, $Z(\alpha)$, but then an
additional step is required to determine the projected arc length
parameter $\xi$. This is easily done using the integral formula
\begin{gather}
  \xi(\alpha) = \int_0^\alpha \sqrt{X_\alpha^2 + Y_\alpha^2} \, d\alpha
  \label{eq:xi-arclength}
\end{gather}
to compute point values of $\xi_i=\xi(\alpha_i)$, which can then be 
interpolated with a spline $\Xi(\alpha)$.

Another quantity of primary importance in skiing is the distance along
the snow surface, denoted $s$, which can also be expressed as an arc
length integral:
\begin{gather}
  s(\alpha) = 
  \begin{cases}
    \ds\int_0^\alpha \sqrt{\Xi_\alpha^2 + Z_\alpha^2}\; d\alpha ,
    & \text{in 2D,}\\[0.4cm]
    \ds\int_0^\alpha \sqrt{X_\alpha^2 + Y_\alpha^2 + Z_\alpha^2}\; d\alpha ,
    & \text{in 3D.}
    \end{cases}
  \label{eq:s-arclength}
\end{gather}
To be consistent with other quantities used to represent course
geometry, $s$ is also interpolated with a spline to obtain
$S(\alpha)$\footnote{If skier position is parameterized by time as
  $\vec{r}(t)$, then arc length may be expressed elegantly as
  $s(t)=\int_0^t \left\| \vec{r}^{\,\prime}(t) \right\| dt = \int_0^t
  |v(t)|\, dt$, where $v(t)$ is the skier speed. However, we will not
  use this formula in our \Matlab\ implementation.}.

When implementing these distance calculations in \Matlab, the integrals
in~\eqref{eq:xi-arclength} and~\eqref{eq:s-arclength} must be computed
numerically. To obtain accurate approximations for $s$ (and also $\xi$
for 3D courses), the coordinate splines ($\Xi, Z$ in 2D, $X,Y,Z$ in 3D)
are first evaluated on a fine grid of $\alpha$ points. For the 4.2~km
Ole course, we use 840 points which yields an average resolution of 5~m
for each $\alpha$-interval.  The integrals are then approximated using
Simpson's rule, which is fourth-order accurate and has a particularly
simple implementation when the $\alpha$ points are equally spaced (see
the \Matlab\ code {\tt cumsimpson.m} provided with the Supplementary
Materials\footnote{Conveniently, the spline data structure used in
  \Matlab\ permits spline derivatives $X_\alpha$, $Y_\alpha$, $Z_\alpha$
  appearing in the arc length integrands to be computed exactly with the
  function {\tt fnder}. We initially attempted to use a finite
    difference approximation for these derivatives, but this generated
    errors that interfered with the convergence of the ODE
    solver.}). Simpson's rule is preferable to \Matlab's built-in
trapezoidal rule function {\tt trapz} because of its much higher
accuracy. Once the corresponding fine-grid values of $\xi$ are computed,
we can then re-parameterize the curve in terms of $\xi$ by computing a
new spline for $Z(\xi)$ (plus $X(\xi)$ and $Y(\xi)$ in 3D). Indeed, the
smooth elevation curve plotted in Fig.~\ref{fig:FISOle}b was generated
by first approximating the integral \eqref{eq:xi-arclength} with
Simpson's rule, and then fitting a Hermite spline to the resulting
$\xi, z$ data.

One more geometric quantity that is of critical importance for capturing
skier dynamics is the angle of inclination $\inclination(\xi)$ that the
course makes with the horizontal at any location $\xi$ (refer to
Fig.~\ref{fig:theta-2D}).  This angle can be expressed in terms of
$z(\xi)$ by recognizing that it is related to the slope
$\frac{dz}{d\xi}$ by
\begin{gather}
  \inclination(\xi) = \arctan\left(\frac{dz}{d\xi} \right) . 
  \label{eq:inclination} 
\end{gather}
Because we have already built the spline $Z(\xi)$ for elevation, the
derivative $Z_\xi$ can be computed exactly using the {\tt fnder}
function from \Matlab's Curve Fitting Toolbox.  Then $\inclination$ can
be evaluated at spline points and used to construct another spline
$\Theta(\xi)$ for the inclination angle.  We have implemented the
procedure described above in the \Matlab\ code {\tt setup2d.m}, which
can be summarized in four main steps:
\begin{itemize}
\item Read the course $\xi,z$ coordinates from the CSV file.
\item Build the spline interpolant $Z(\xi)$ (the default spline type is
  {\tt 'makima'} but can be modified).
\item Compute the spline derivative $Z_\xi$ and use it to construct
  splines $S(\xi)$ and $\Theta(\xi)$.
\item Print diagnostics regarding point spacing, arclength and 
  course gradient for homologation purposes.
\end{itemize}
\noindent
More details are provided in Section~\ref{sec:3Dmodel} regarding the
corresponding code {\tt setup3d.m} for constructing a 3D parametric
spline representation from GPS data.

In closing, we reiterate that in~\cite{moxnes2014using}, MSH based the
majority of their simulations on a linear spline representation of the
course. They compared their results to a single run with a cubic spline
interpolant, yielding a finish time roughly 20~s slower, which is a
reflection of the fact that a linear spline always yields the shortest
possible distance between data points. In their view, the cubic spline
result was close but still yielded a poorer match with the measured race
time forming the basis for their comparisons. This led them to conclude
that their model simulations with the cubic spline were ``not more
realistic than the chosen piecewise linear track,'' despite the fact
that any differences could just as easily be attributed to their choice
of model parameters.  We should also recognize that piecewise linear
interpolation of GPS data is inherent in the ``homologation'' procedure
that FIS officials use to measure and certify Nordic race courses (and
which is discussed further in the next section). One of our main goals
is therefore to investigate how employing a smooth spline interpolant
affects the accuracy of model results in comparison with a linear
approximation.

\begin{exercise}
  \label{ex:dzdx}
  The model devloped by MSH~\cite{moxnes2014using} parameterizes a ski
  course as $z(s)$ using arc length $s$, and replaces slope
  $\frac{dz}{d\xi}$ in Eq.~\eqref{eq:inclination} with
  $\frac{dz}{ds} \big/ \sqrt{1 - (\frac{dz}{ds})^2}$.  Show that these
  two expressions for slope are equivalent for a ``reasonable'' ski
  course with no self-intersections and finite slope.
\end{exercise}

\begin{exercise}
  \label{ex:ole-quintic}
  Interpolate the Ole course data with a quintic (degree 5) polynomial. 
  This can be done using the \Matlab\ function {\tt spapi} instead 
  of {\tt makima}.  Plot the quintic spline 
  alongside the cubic Hermite interpolant and discuss any ``problematic'' 
  features of the quintic approximation.  Because {\tt spapi} and {\tt makima} 
  return different spline data structures, generating plots is easiest with the 
  all-purpose {\tt fnplt} function.
\end{exercise}

%%%%%%%%%%%%%%%%%%%%%%%%%%%%%%%%%%%%%%%%%%%%%%%%%%%%%%%%%%%%%%%%%%%%%%
\subsection{Homologation and Data Errors}
\label{sec:homologation}

All ski courses that run high-level Nordic races must adhere to
guidelines set out in the \emph{FIS Cross-Country Homologation
  Manual}~\cite{FISHomologation-2021}.
% According to the dictionary, ``homologation'' is the granting
% of approval by some official authority.
In the context of skiing, homologation refers to
the FIS-certified process that any race course must go through before an
official competition can be held. The Manual sets out constraints on
course terrain and stadium layout, including limits on the number,
length, and slope of hills that are allowed\footnote{An
  easily-digestible summary of the homologation rules can be found on
  the
  \href{https://en.wikipedia.org/wiki/Cross-country_skiing_trail\#Competition}
  {Wikipedia page ``Cross-country skiing trail''}.}.  These
specifications are implemented in the form of an Excel spreadsheet that
FIS officials use to perform the necessary
calculations~\cite{FISHomologation-2021}.

A basic understanding of these course design criteria is useful,
especially in light of the reality that any GPS measurements must
contain errors, which could cause a spline interpolant to deviate
significantly from the certified course.  Below is a list of the main
homologation criteria, which we number as H1--H6:
\begin{enumerate}[label=H\arabic*.]
\item GPS measurements should be taken every 20~m along a course, or at 
  every point where the gradient changes. % p.50
  Regardless of the GPS resolution, our fine-grid of 5~m for spline calculations 
  was chosen to fall well within this 20~m distance constraint. If we 
  interpret ``gradient changes'' as points where the slope changes sign, then 
  this is a further justification for using Hermite spline interpolation 
  since it avoids introducing any new points where the gradient might change.
  
\item The actual length of a race course should not exceed the
  officially posted distance by more than 10\%, nor should it be shorter
  by more than 5\%.
  
\item The average gradient, including all uphills and downhills, should
  be in the range of 6--14\%.
  % \footnote{For comparison, the FIS rules for alpine races mandate
  % gradients between 33--45\%.}.   
  This corresponds to an inclination angle $\inclination$ (see
  Fig.~\ref{fig:theta-2D}) lying between 3.4${}^\circ$--8.0${}^\circ$,
  since percentage gradients refer to a fraction of
  45~degrees\footnote{To convert between percentage gradients and
    angles, use the fact that the instantaneous slope of a curve
    $z(\xi)$ is the derivative $\frac{dz}{d\xi}$. Then the percent
    gradient is $\text{\emph{pct}} = 100 \frac{dz}{d\xi}$, which gives
    these two conversion formulas:
    $\text{\emph{pct}} = 100 \tan\left(\frac{\pi\inclination}{180}
    \right)$ and
    $\inclination = \frac{180}{\pi} \, \arctan\left(
      \frac{\text{\emph{pct}}}{100} \right)$.  A nice diagram relating
    $\inclination$ and \emph{pct} can be found on the
    \href{https://en.wikipedia.org/wiki/Grade_(slope)}{Wikipedia page
      ``Grade\ (slope)''}.}.

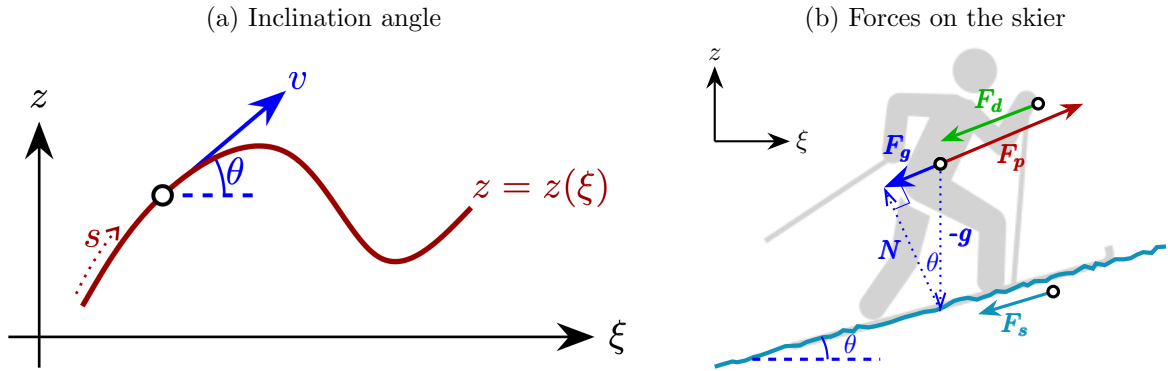
\begin{figure}[tb]
  \centering
  \begin{tabular}{ccc}
    (a) Inclination angle && (b) Forces on the skier \\
    \begin{tikzpicture}[scale=0.65]
      \node[above right,inner sep=0pt,scale=1.15,red!70!black,] at (7,2.6) {$z=z(\xi)$};
      \draw[blue,very thick,dashed] (2,2.8)--(3.5,2.8);
      \draw[blue,-{Stealth[length=4mm]},very thick] (2,2.8)--(4,4.5) node[above right,blue,scale=1.15,inner sep=0pt] {$v$};
      \draw[blue,thick] (3,2.8) arc (0:30:1.3) node[right,scale=1.15,inner sep=0pt] at (3,3.2) {$\inclination$};
      \draw[red!70!black, ultra thick] (0.7,1) to[spline through={(2,2.8) (4,3.5) (5.5,1.8)}] (7,2.6);
      \draw[black, very thick, fill=white] (2,2.8) circle (1.5mm);
      \draw[thick,-{Stealth[length=4mm]}] (-0.5,0.5)--(9,0.5) node[right,scale=1.15] {$\xi$};
      \draw[thick,-{Stealth[length=4mm]}] (0,0)--(0,4) node[above,scale=1.15] {$z$};
      \draw[red!70!black,thick,dotted,-{Straight Barb[width=1.5mm]}] (0.6,1.2) arc (155:140:5) node[left,scale=1.15,inner sep=1pt] at (1.1,2.1) {$s$}; 
    \end{tikzpicture}
  &\qquad&
  \begin{tikzpicture}[scale=1]
    % Skier diagram
    \node[inner sep=0pt] (skier) at (1,0.72) 
    {\includegraphics[width=0.3\textwidth]{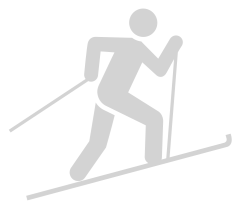}};
    % Snow surface and inclination angle
    % \draw[ultra thick] (-2,-1.5) -- (4,0.1);
    \draw[ultra thick,cyan!70!black,decorate, decoration={random steps,segment length=3pt,amplitude=1pt}] (-2,-1.5) -- (4,0.1);    
    \draw[very thick,dashed,blue] (-1.5,-1.4) -- (0.2,-1.4);
    \draw[blue,thick] (-0.5,-1.4) arc (0:30:0.6) node[right,inner sep=0pt] at (-0.3,-1.2) {$\inclination$};
    % Coordinate axes
    \draw[thick,-{Stealth[length=3mm]}] (-2,1.5) -- (-1,1.5) node[right,scale=1,inner sep=2pt] {$\xi$}; 
    \draw[thick,-{Stealth[length=3mm]}] (-2,1.5) -- (-2,2.5) node[above,scale=1,inner sep=2pt] {$z$};
    % Force arrows
    \draw[thick,dotted,blue,-{Straight Barb[length=2mm,width=1.3mm]}] (1,1.2) -- (1,-0.76) node[midway,right] {$\pmb{\text{-}g}$};
    \draw[very thick,blue,-{Stealth[length=4mm]}] (1,1.2) -- (0.25,0.8875) node[near end,above,inner sep=7pt] {$\pmb{F_g}$};
    \draw[thick,dotted,blue,-{Straight Barb[length=2mm,width=1.3mm]}] (1,-0.76) -- (0.25,0.8875) node[midway,left] {$\pmb{N}$};
    \draw[black,fill=white,very thick] (1,1.2) circle (0.7mm);
    \draw[very thick,red!70!black,-{Stealth[length=3mm]}] (1,1.2) -- (2.9,2) node[midway,below] {$\pmb{F_p}$};
    \draw[very thick,green!70!black,-{Stealth[length=3mm]}] (2.3,2)  --  (1,1.5) node[midway,above] {$\pmb{F_d}$};
    \draw[very thick,cyan!70!black,-{Stealth[length=3mm]}] (2.5,-0.5)  --  (1.5,-0.8) node[midway,below] {$\pmb{F_s}$};
    \draw[black,fill=white,very thick] (1,1.2) circle (0.7mm);
    \draw[black,fill=white,very thick] (2.3,2) circle (0.7mm);
    \draw[black,fill=white,very thick] (2.5,-0.5) circle (0.7mm);
    \draw[blue] (0.9,-0.1) node {$\inclination$};
    \draw[blue,thin] (0.45,0.97) -- (0.59,0.68) -- (0.4,0.6);
  \end{tikzpicture} 
  \end{tabular}
  \caption{(a) A 2D elevation profile with the course parameterized as
    $z=z(\xi)$, showing the inclination angle $\inclination$, arc length
    $s$ (or skier distance), and speed $v$. (b) A skier is acted upon by
    forces due to propulsion ($F_p$), gravity ($F_g$), drag ($F_d$) and
    snow friction ($F_s$).}
  \label{fig:theta-2D}
\end{figure}

\item All courses should be roughly balanced between uphills, downhills,
  and undulating terrain. Hills are classified as having gradients
  between 9--18\%, corresponding to inclination angles of
  5.1${}^\circ$--10.2${}^\circ$, and are further divided by length into
  short hills (extending for 10--29~m) and long hills (over
  30~m). Gradients above 18\% are allowed, but only over distances of
  10~m or less.
  
\item The race start and finish should both consist of a straight
  section roughly 100--150~m long, with the finish having a gentle
  upward climb with slope 2--4\%.

\item To ensure athlete safety, a maximum limit on curvature of downhill
  sections is imposed in terms of centripetal acceleration, $a_c=v^2/R$,
  where $v$ is the skier speed and $R$ is the radius of curvature. A
  table in \cite[p.~26]{FISHomologation-2021} identifies various limits
  on $a_c$ ranging from 2--25~m/s${}^2$ depending on race format.
\end{enumerate}
These design criteria will be discussed further after we have developed
an algorithm for approximating the course arc length, slope, and
curvature in the following sections. Once we have investigated whether
any homologation rules are violated, we will consider whether (and how)
the input data should be modified to more accurately capture a realistic
race course.

\begin{exercise}
  \label{ex:start-missing}
  Notice in Fig.~\ref{fig:FISOle}b that the Ole course begins
  immediately with an uphill gradient, which apparently violates
  homologation criterion H5. This suggests that perhaps the start area
  was omitted when GPS measurements were taken.  With this in mind,
  modify {\tt setup3d.m} to extend the start with a straight, flat
  section of length 100~m, defined by adding two extra points at the
  same elevation as the original start. To determine $x,y$ components
  for these points, use linear extrapolation based on the first two GPS
  points.  Plot the extended course as a 2D elevation profile.
\end{exercise}

\begin{exercise}
  \label{ex:gps-repeated}
  Download the GPS file for either the ``Altprags Uphill'' or
  ``Nathalie'' course from the Do\-lo\-mi\-ti Nordic\-Ski
  website~\cite{dolomiti-gps-2023}. Verify that at least one pair of GPS
  points are duplicates (that is, they have identical coordinates
  $\latitude$, $\longitude$, $z$).  When this occurs the $\xi_i$ are not
  monotone increasing, which can cause problems!  Modify {\tt setup3d.m}
  to automatically identify and delete any duplicate points before
  constructing the spline interpolant.
\end{exercise}

\begin{exercise}
  \label{ex:homologate}
  Download the GPS file for one of the FIS-rated courses named
  ``Stephanie'' or ``Saskia'' from the Dolomiti NordicSki website
  \cite{dolomiti-gps-2023}. Run the {\tt setup3d.m} code with this data
  and investigate whether the resulting spline satisfies the
  homologation criteria H1--H5 (don't worry about curvature and H6 for
  now).
\end{exercise}

%%%%%%%%%%%%%%%%%%%%%%%%%%%%%%%%%%%%%%%%%%%%%%%%%%%%%%%%%%%%%%%%%%%%%%
% --------------------------------------------------------------------
%%%%%%%%%%%%%%%%%%%%%%%%%%%%%%%%%%%%%%%%%%%%%%%%%%%%%%%%%%%%%%%%%%%%%% 

\section{2D Model of Nordic Skiing}
\label{sec:2Dmodel}

\subsection{ODE System for Skier Dynamics}
\label{sec:2Dmodel-ODEs}

We begin by developing a model for skiing along a 2D course where
elevation is a function of distance $z(\xi)$, as shown in
Fig.~\ref{fig:FISOle}b. There are several reasons for this choice:
\begin{itemize}
\item The course parameterization is simpler, requiring only a single
  spline curve, and the governing ODEs are also simpler.
\item Most existing dynamic models of Nordic
  skiing~\cite{carlsson-etal-2011, moxnes2008cross, moxnes2014using,
    sundstrom2013numerical}
  % Ni et al, 20222 is in 3D 
  are also two-dimensional, with the skier assumed to move along the 2D
  course elevation profile.
\item Starting in 2D avoids the complexity of curvature and braking,
  allowing us to concentrate on testing and calibrating our results
  against other models before taking the next step to 3D.
\end{itemize}
Our aim is therefore to determine the skier position
$\vec{r}(t) = \big(\xi(t),\, z(t)\big)$ as a function of time, with $t$
measured in seconds.  The skier speed (in m/s) is given by
$\vec{v}(t) = \vec{r}^{\,\prime}(t) = \big(\xi^\prime(t),\, z^\prime(t)
\big)$, where the ``prime'' denotes a time derivative, and the scalar
speed in the direction of motion is the vector magnitude
$v(t) = \|\vec{v}(t)\| = \sqrt{\xi^\prime(t)^2 + z^\prime(t)^2}$.  This
requires that speed is always positive ($v>0$), meaning that the skier
always moves in the forward direction, which is a reasonable assumption
for trained athletes.

The skier is treated as a point mass that moves along the curve defining
the course, and is assumed to obey Newton's second law -- ``mass times
acceleration equals force.'' The forces acting on the skier are defined
for convenience as forces per unit mass (with units of m/s$^2$) and are
assumed to come from four main sources:
\begin{itemize}
\item \emph{Propulsion force $F_p(v)$}: which is generated internally by
  the skier's muscles and incorporates various effects such as aerobic
  and anaerobic metabolism, skiing style (skate or~classic), technique
  (diagonal stride, double pole, offset, one-skate, two-skate, etc.) and
  possibly extra effects arising from technique or race strategy. We use
  the following speed-dependent function
  \begin{gather}
    F_p(v) = \frac{P(v)}{mv},
    \label{eq:Fp}
  \end{gather}
  where $m$ is the skier mass (in kg) and $P(v)$ represents the power
  (in watts or kg\,m${}^2$/s${}^3$) that is exerted by the muscles to
  generate forward motion. The ``$\frac{1}{v}$'' dependence implies that
  the skier works hardest to accelerate at low speeds, whereas the power
  demands on their muscles ease off at higher speeds\footnote{Our
    earlier assumption that $v>0$ ensures that this propulsion force is
    well-defined. However, we could handle a zero speed by replacing the
    factor $v$ in the denominator of~\eqref{eq:Fp} with
    $\max(v,\varepsilon)$ for some small cut-off parameter
    $\varepsilon>0$.}.  This function was one of three different
  functional forms proposed by MSH~\cite{moxnes2014using} and fit to
  power measurements of Nordic ski racers, and the specific form for
  $P(v)$ will be described later in Section~\ref{sec:2Dmodel-params}.
    
\item \emph{Gravitational force
    $F_g(\inclination) = -g\sin\inclination$:} where $g=9.81$~m/s${}^2$
  is the acceleration due to gravity. Note that $F_g$ is positive on
  downhill sections where $\inclination<0$ (acting to speed up the
  skier), whereas it is negative on uphills where $\inclination>0$
  (slowing the skier down).
  
\item \emph{Snow friction
    $F_s(\inclination) = -\mu g \cos\inclination$:} where $\mu$ is the
  (dimensionless) coefficient of dynamic friction. According
  to~\cite{carlsson-chapter-2016, moxnes2014using}, typical values of
  $\mu$ lie in the range $[0.03,0.06]$, although in general it depends
  on a variety of factors including air/snow temperature, sun exposure,
  grooming quality, ski wax, etc.  Note that friction always opposes
  motion whether the skier is travelling up/downhill.

\item \emph{Aerodynamic drag force $F_d(v) = - \beta v^2$:}
  % \widetilde{v} |\widetilde{v}|
  which we assume depends quadratically on the speed relative to still
  air\footnote{If there is a significant headwind/tailwind, then $v$
    should be replaced with a wind-corrected speed (see
    Exercise~\ref{ex:wind}).}. The parameter
  $\ds\beta=\frac{\rho C_d A}{2m}$ (units of m$^{-1}$) is proportional
  to air density $\rho$ (in kg/m${}^3$), skier cross-sectional area $A$
  (m${}^2$), and a dimensionless drag coefficient $C_d$. As their speed
  varies throughout a race, trained skiers will adjust their technique
  and body orientation so that the area $A$ varies with the skier's
  speed. So in practice, we will consider the product $C_dA(v)$ to be a
  given function of speed (discussed in more detail in
  Section~\ref{sec:2Dmodel-params}).
\end{itemize}
Newton's second law then requires that the sum of all applied forces
must balance the acceleration:
\begin{gather*}
  m v^\prime = m \Big[ F_p(v) + F_g(\inclination) + F_s(\inclination) + F_d(v) \Big].
\end{gather*}
It should now be clear why we chose to define forces per unit mass,
since this permits us to eliminate the skier mass $m$ from all terms
except the propulsion force and obtain the simplified equation
\begin{align}
  v^\prime &= \frac{P(v)}{mv} - g\sin\inclination - 
  \mu g \cos\inclination  - \beta v^2. \label{eq:ode2a}
\end{align}
This equation is an ODE for $v$ that depends implicitly on the course
geometry through the inclination angle as $\theta=\theta(\xi(t))$, so we
still need another equation for the time evolution of $\xi$.  The time
derivative of $\xi$ is just the horizontal component of the skier
velocity (see Fig.~\ref{fig:theta-2D}), which can be written as the ODE
\begin{align}
  \xi^\prime &= v\cos\inclination. \label{eq:ode2b}
\end{align}
The inclination angle appearing in the right-hand side of both ODEs can
be obtained by simply evaluating the spline $\theta=\Theta(\xi)$
constructed in Section~\ref{sec:distances}.

This completes the model for a 2D skier, which can be summarized
compactly as follows. The skier speed $v(t)$ and projected arc length
$\xi(t)$ are integrated in time by solving the two ODEs
\eqref{eq:ode2a}--\eqref{eq:ode2b}.  Wherever $\theta$ appears in the
equations, it is replaced with the corresponding spline $\Theta(\xi(t))$
evaluated at the current value of $\xi$.  Furthermore, the elevation
$z(t)$ and skier distance $s(t)$ can be determined at any time by
evaluating the splines $Z(\xi(t))$ and $S(\xi(t))$ respectively.  This
is an especially simple and efficient algorithm because all splines are
derived from the given course geometry, and so can be precomputed before
solving any ODEs. There is an additional advantage that regardless of
any numerical errors introduced in the calculation of $v(t)$ and
$\xi(t)$, the skier's position will always lie exactly on the spline
used to construct the course.

To obtain a well-posed initial value problem, the above ODEs must be
supplemented with initial conditions. Since we shifted the course
coordinates so that the start lies at the origin, it is natural to
initialize the projected distance to $\xi(0)=0$.  For the speed, we take
the same initial value $v(0)=2$ used by MSH\footnote{A more natural
  choice of initial condition might be $v(0)=0$.  However, we can never
  allow the skier to be at rest (or moving at very low speed) because
  the propulsion force has $F_p(v)\to\infty$ as $v\to 0$. This is a
  consequence of MSH fitting their power function $P(v)$ to measurements
  of athletes who were already in motion. Our choice of $v(0)=2$~m/s is
  reasonable as long as we recognize that $t=0$ actually corresponds to
  a few seconds after the race start, once the athlete has gotten up to
  speed.  In practice, taking $v(0)$ much smaller than 2 can cause
  anomalous behaviour owing to unrealistically large propulsion
  forces.}.

\begin{exercise}
  \label{ex:dimanal-check}
  Verify that the ODE system~\eqref{eq:ode2a}--\eqref{eq:ode2b} is
  ``dimensionally consistent'' in the sense that within each equation,
  all terms have the same physical units.
\end{exercise}

\begin{exercise}
  \label{ex:dimanal-Fprop}
  Use dimensional analysis (by applying the Buckingham Pi
  Theorem~\cite{santiago-2019}) to derive a general form for the
  propulsion force in Eq.~\eqref{eq:Fp}, assuming only that $F_p$
  depends on the quantities $\Pmax$, $m$ and $v$.
\end{exercise}

\begin{exercise}
  \label{ex:exact2D-hat}
  Derive an analytical solution for the ODE~\eqref{eq:ode2a} on a linear
  course with constant slope.
  \begin{enumerate}[label=(\alph*)]
  \item Find an explicit formula for speed $v(t)$ when the course
    profile is $z(\xi) = b\xi$ for some constant slope $b$. Assume that
    the skier applies a constant propulsion force,
    $F_p(v) \equiv F_{po}$, which implies that the MSH power function is
    linear in the speed, $P(v) \propto v$. In constrast with
    Eq.~\eqref{eq:Fp}, this force has no ``$\frac{1}{v}$'' singularity,
    so you may also assume that the skier starts from rest with
    $v(0)=0$.
  \item Plot the speed $v(t)$ for an uphill climb with a gradient of
    $+$10\% ($b=0.1$) and assume parameter values $\beta=0.0045$,
    $g=9.8$, $\mu=0.037$, and $F_{po}=1.4$ from the MSH baseline test
    (see Table~\ref{tab:params} and Fig.~\ref{fig:moxnes-sims}). Observe
    that the skier's speed initially increases but then gradually
    flattens out over time and approaches a constant terminal value.
    Estimate this terminal speed and the time required to reach it.
  \item Repeat part (b) on a constant downhill gradient of $-$10\%,
    assuming the skier rests in a tuck position with $\beta=0.0025$ and
    $F_{po}=0$. Compare your results for the uphill and downhill slopes.
  \end{enumerate}
\end{exercise}

%%%%%%%%%%%%%%%%%%%%%%%%%%%%%%%%%%%%%%%%%%%%%%%%%%%%%%%%%%%%%%%%%%%%%%
\subsection{Augmented ODE System}
\label{sec:alt-ODEs}

A closely-related model for skier dynamics in~\cite{carlsson-etal-2011,
  sundstrom2013numerical} treats $z(t)$ and $s(t)$ as additional
solution variables and derives ODEs for them based on the course
geometry. The time evolution of $z$ and $s$ is determined solely by the
components of velocity in the vertical and tangential directions at any
point along the course, leading to these two simple ODEs
\begin{align}
  z^\prime &= v\sin\inclination, \label{eq:ode2c}\\
  s^\prime &= v, \label{eq:ode2d}
\end{align}
which are supplemented by initial conditions $z(0)=s(0)=0$.  Altogether
the augmented system~\eqref{eq:ode2a}--\eqref{eq:ode2d} consists of
4~nonlinear ODEs in the unknowns $v(t)$, $\xi(t)$, $z(t)$ and $s(t)$,
where the inclination angle is obtained by a spline evaluation
$\Theta(\xi(t))$.  In contrast with the two-ODE model from the previous
section, the elevation $z(t)$ is no longer constrained to lie exactly
along the skier path because numerical errors in the solution
of~\eqref{eq:ode2c} cause the skier location to ``drift.'' Unless these
errors are controlled, they can generate significant deviations in the
skier location away from the course, with a similar drift occurring in
$s(t)$.

%%%%%%%%%%%%%%%%%%%%%%%%%%%%%%%%%%%%%%%%%%%%%%%%%%%%%%%%%%%%%%%%%%%%%%
\subsection{Fitting Parameters to Skier Measurements}
\label{sec:2Dmodel-params}

Unless indicated otherwise, all of our numerical simulations are
performed using ``baseline'' parameters taken from
MSH~\cite{moxnes2014using} that are listed in Table~\ref{tab:params}. In
particular, their baseline test subject for both experiments and
simulations was a male skier with mass $m=77.5$~kg. The gravitational
acceleration is $g=9.81$~m/s${}^2$, while the air density is
$\rho = 1.29$~kg/m${}^3$.

\begin{table}[bth]
  \centering
  \caption{Parameter values for the baseline test 
    case in MSH~\cite{moxnes2014using}, reported in SI units.}
  \renewcommand{\arraystretch}{1.2}
  \begin{tabular}{lccc}
    Parameter & Symbol & Value or Range & Units \\\hline
    Gravitational acceleration & $g$ & 9.81 & m/s${}^2$ \\
    Air density & $\rho$ & 1.29 & kg/m${}^3$ \\
    Skier mass & $m$ & 77.5 & kg \\
    Snow friction coefficient & $\mu$ & 0.037  & -- \\
    Drag coefficient & $C_dA(v)$ & Eq.~\eqref{eq:CdA} & m${}^{2}$\\
    Drag parameter & $\beta=\ds\frac{\rho}{2m} \, C_dA(v)$ & $[0.0029,0.0046]$ & m${}^{-1}$\\
    Maximum power & $\Pmax$ & 275 & kg\,m${}^2$\,s${}^{-3}$\\
    Power parameter & $b$ & 10 & m/s \\
    Power exponent & $n$ & 4 & --\\
    % Braking coefficient (skid, step) & $\gamma$ & 0.0026, 0.0058 &  --\\
    \hline
  \end{tabular}
  \label{tab:params}
\end{table}

The drag coefficient $C_d A$ varies throughout a race because athletes
adjust their posture between an upright or tuck position that alters
their frontal cross-sectional area $A$. At the two extremes, a skier
moving at high speed on a steep downhill will fall into a deep tuck
position to minimize drag and conserve energy; however, at lower speeds
(especially when skiing uphill) the skier must focus more on generating
forward motion which requires that they adopt a more upright posture.
It then seems reasonable to assume the drag coefficient is a function of
speed, so we follow MSH and take $C_dA(v)$ to be a piecewise constant
function of speed that switches at a given threshold $v\gtrless 10$~m/s
between two drag values:
\begin{gather}
  C_d A(v) =
  \begin{cases} 
    0.55, & v \leqslant 10~~~~\text{(upright)}, \\
    0.35, & v > 10~~~~\text{(tuck)}.
  \end{cases}
  \label{eq:CdA}
\end{gather}

The final remaining detail is the locomotive power function $P(v)$
introduced in Eq.~\eqref{eq:Fp}. MSH used extensive measurements of ski
racers to propose three different functional forms for $P(v)$, which
they compared using numerical simulations. We will adopt an exponential
function which they refer to as ``Model~1'' and used for their baseline
simulations\footnote{The power functions corresponding to MSH Models 2,3
  have a similar shape to the baseline, and don't have a major impact on
  the resulting solution, something the interested reader is welcome to
  explore.}:
\begin{gather}
  P(v) = \Pmax \exp\big(\text{--}\,(v/b)^n \big). 
  \label{eq:Pv}
\end{gather}
This function is designed to mimic an athlete who generates the highest
power when skiing uphill (at lower speeds) and the lowest on downhills
(at high speeds). This reflects the same adjustments in technique we
discussed for the drag coefficient, where the skier has to use their
entire body to work hardest on steep uphill sections and maintain their
speed, whereas on steep downhills they shift into a tuck position to
conserve energy and minimize drag. An intermediate level of power is
expended on undulating terrain, where a relatively high speed can be
maintained using less energy-intensive techniques.  MSH estimated the
parameters in this power equation by fitting with experimental data to
obtain $n = 4$, $b = 10$, and $\Pmax = 275$.  These authors indicate
that their maximum power value was obtained from measurements of an
elite athlete who was ``skiing at a self-chosen pace.''  We will see
later in comparison with another set of experiments by Welde on elite
racers (Section~\ref{sec:2Dsims-Welde}) that even though this athlete's
pacing might be considered fast, they were almost surely not skiing at a
race pace consistent with elite skiers.

The next section presents several examples of simulations that start
from skiers with the baseline parameter values, and explore how changes
in certain parameters impact the solution. A more comprehensive
parameter sensitivity study can be found in MSH.

%%%%%%%%%%%%%%%%%%%%%%%%%%%%%%%%%%%%%%%%%%%%%%%%%%%%%%%%%%%%%%%%%%%%%% 
% --------------------------------------------------------------------
%%%%%%%%%%%%%%%%%%%%%%%%%%%%%%%%%%%%%%%%%%%%%%%%%%%%%%%%%%%%%%%%%%%%%% 
\section{2D Model Simulations}
\label{sec:2Dsims}

The system of nonlinear ODEs \eqref{eq:ode2a}--\eqref{eq:ode2b}, or its
augmented formulation \eqref{eq:ode2a}--\eqref{eq:ode2d}, can be solved
easily and accurately using \Matlab's variable-step, variable- order ODE
solver {\tt ode113}. This solver employs Adams-Bashforth-Moulton
formulas with derivative approximations having orders from 1 to 13,
adaptively selecting both the order and the time step so as to minimize
computational cost while also satisfying the given error tolerances. A
comprehensive comparison of this and other \Matlab\ ODE solvers can be
found in the paper~\cite{shampine-reichelt-1997}, and
Exercise~\ref{ex:ODEsolvers} provides a chance to compare the
performance of several different solvers.

This approach has been implemented in our code {\tt skirun2d.m} with the
parameter {\tt neqs} (the number of ODEs) set to either 2 or 4 to select
the desired problem formulation.  In the following sections, we apply
the code to solving three test examples:
\begin{enumerate}
\item The MSH baseline simulation~\cite{moxnes2014using} on a 4.2~km
  course using the reduced (2-ODE) model.  This provides a convenient
  test to validate our \Matlab\ code and investigate any differences
  that arise due to our choice of ODE solver and spline interpolant
  (linear or cubic Hermite).

\item Comparing to the experimental data in Welde~\cite{welde2017pacing}
  for skiers on a much longer 15~km course, which provides a further
  validation for a different course geometry.

\item The 4.2~km Ole course (introduced in
  Section~\ref{sec:course-spline}) which is a FIS-certified race course
  that is used to further ``stress-test'' the model.
\end{enumerate}
A major advantage of {\tt ode113} is that it adaptively adjusts
the time step and the order of the numerical method to satisfy
user-specified error tolerances (we use {\tt AbsTol = RelTol = 1e-8}).
This ensures that especially in the more difficult sections of the
course (where curvature is high, or at spline junctions where
derivatives lose smoothness) the accuracy of the numerical solution is
maintained, while allowing a much larger time step to be taken in
``easier'' sections.  Furthermore, \Matlab's ODE solvers have built-in
event detection that allows the integration to be terminated precisely
at the end of the course. This is an advantage over models such as
Carlsson et al.~\cite{carlsson-etal-2011}, who employed a constant-step
Runge-Kutta solver, which forced them to choose an unnecessarily small
time step so that they could terminate their code as close as possible
to the finish line.

%%%%%%%%%%%%%%%%%%%%%%%%%%%%%%%%%%%%%%%%%%%%%%%%%%%%%%%%%%%%%%%%%%%%%%
\subsection{Example 1: MSH Baseline Comparison}
\label{sec:2Dsims-MSH}

Here we aim to replicate the results from MSH, who compared experimental
measurements with simulations for a single skate skier on an unnamed
4.2~km course. The course is defined by 50 points with $(s,z)$
coordinates taken from a 2D plot of elevation versus skier
distance~\cite[Fig.~1]{moxnes2014using}. The data are easily obtained
from the given figure using the online data extraction tool
WebPlotDigitizer~\cite{webPlotDigitizer}, by first uploading the plot
image, then calibrating the coordinate axes, manually selecting data
points using mouse-clicks, exporting the $(s_i,z_i)$ coordinates to a
CSV file, and reading the data into \Matlab\ with {\tt csvread}.  The
interested reader is welcome to investigate other approaches for
obtaining course data using a package such as GRABIT (distributed
through the \Matlab\ File Exchange~\cite{grabit-\Matlab-2016}) or the
built-in command {\tt ginput}, which would then permit both the data
extraction and simulation steps to be performed within the \Matlab\
environment.

\begin{figure}[tb]
  \centering
  \begin{tabular}[t!]{cc}
    (a) MSH elevation profile & (b) Simulated speed\\
    \includegraphics[width=0.45\textwidth]{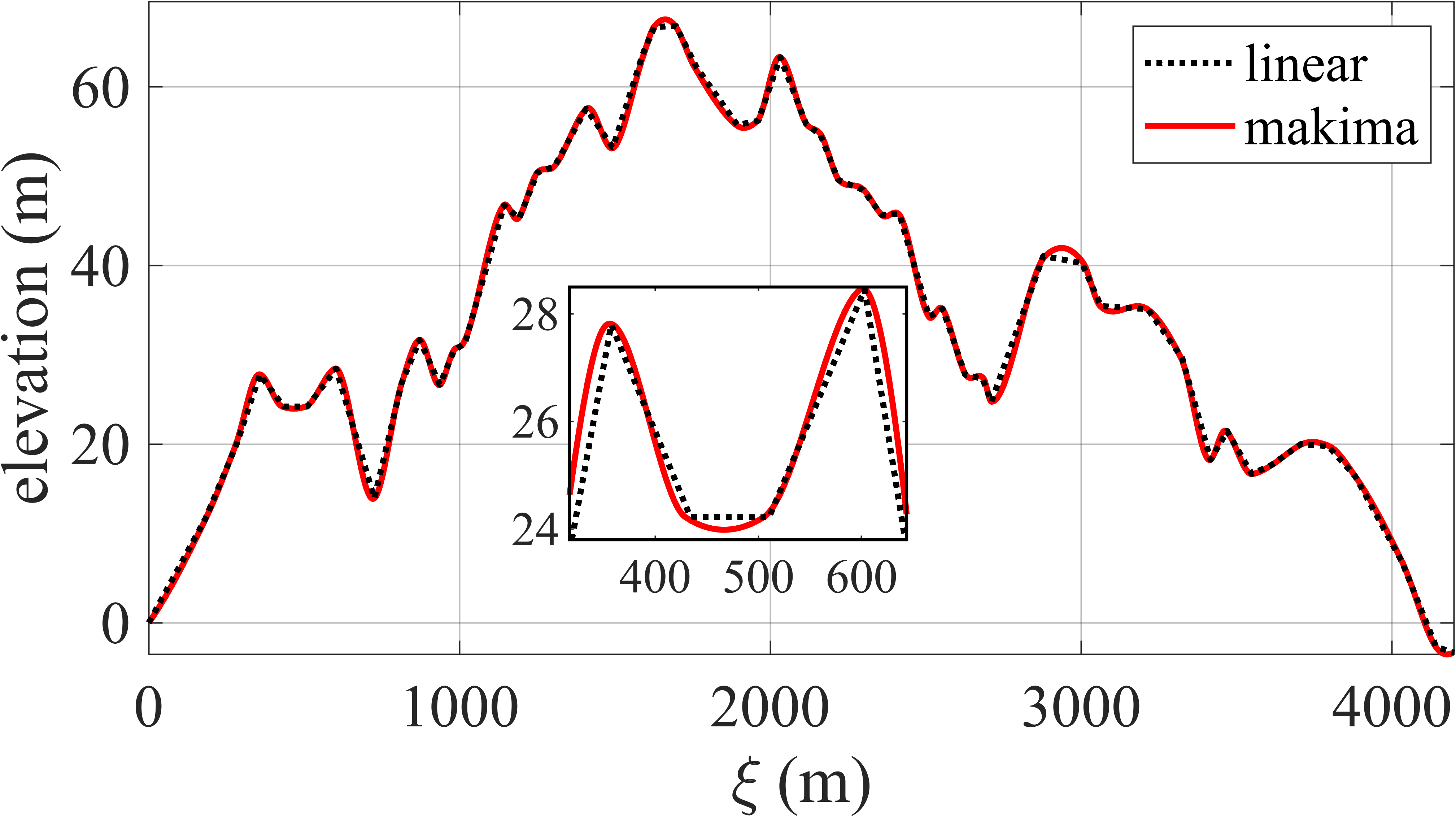}
    & \includegraphics[width=0.45\textwidth]{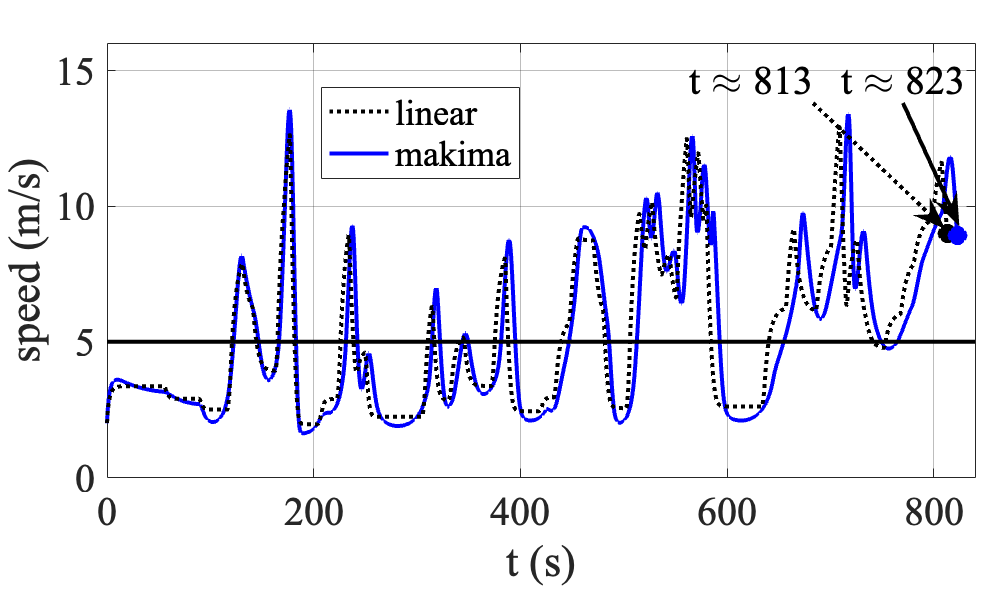}
  \end{tabular}
  \\
  \begin{tabular}{cc}
    (c) Force components & (d) Zoom on a steep uphill\\
    \begin{tikzpicture}
      \node at (0,0) {\includegraphics[width=0.55\textwidth]{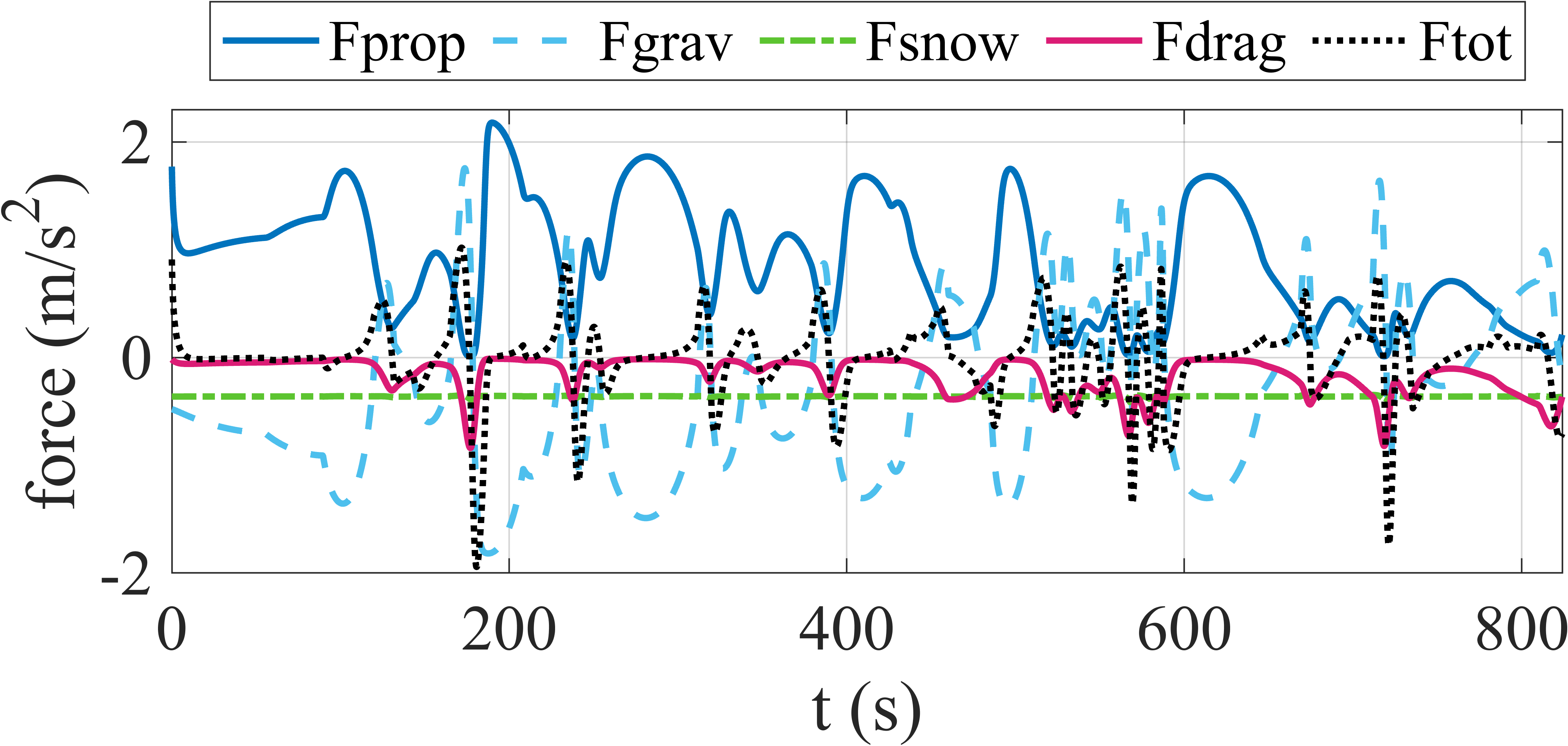}};
      \draw[gray,very thick,dashed] (-1.2,-1.1)--(-0.3,-1.1)--(-0.3,1.45)--(-1.2,1.45)--(-1.2,-1.1);
    \end{tikzpicture}
  & \includegraphics[width=0.25\textwidth]{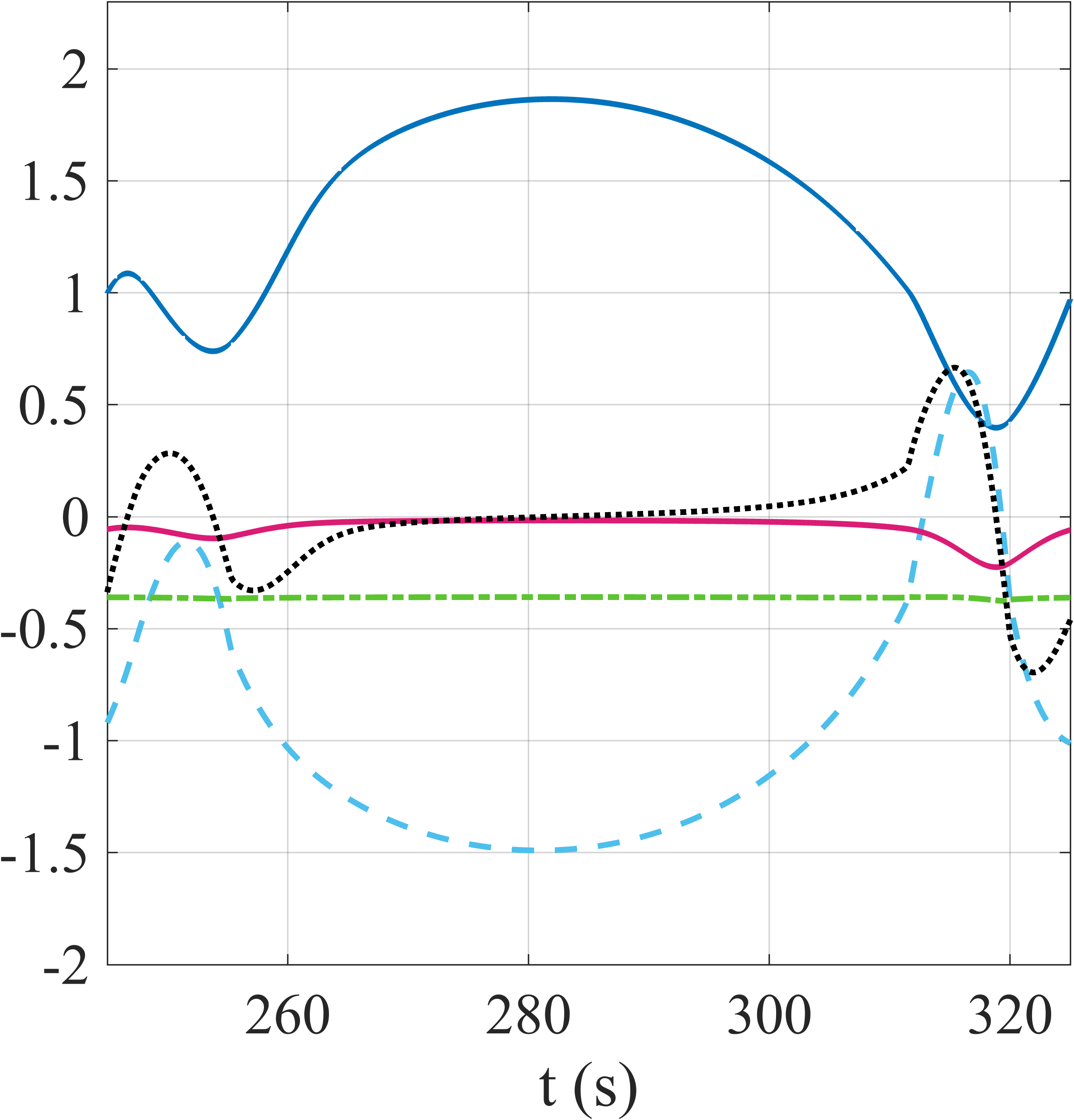}
  \end{tabular}
  % \begin{tabular}[t!]{c}
  %   (e) Elevation and arc length\\
  %   \includegraphics[width=0.45\textwidth]{elevation_arclength_time}
  % \end{tabular}\\
  \caption{Simulations of the MSH baseline case.  (a, Top Left)
    Elevation profile shown in terms of $(\xi,z)$ coordinates (black
    points) extracted from~\cite[Fig.~1]{moxnes2014using}, and
    interpolated with a Hermite (red, solid line) and linear spline
    (black, dotted line). (b, Top Right) Simulated speeds for skiers
    using the two splines, with the average speed of 5.1~m/s shown as a
    horizontal line. (c, Bottom Left) Forces (per unit mass) throughout
    the Hermite spline course, including $F_p$, $F_g$, $F_s$, $F_d$ and
    $\subtext{F}{tot}$. (d, Bottom Right) Zoomed view of the force plot
    on the steep uphill section with $1000\lesssim\xi\lesssim 1150$,
    where the skier accelerates to a roughly constant speed after which
    $\subtext{F}{tot}\approx 0$.}
  \label{fig:moxnes-sims}
\end{figure}

Note that this is a slightly different scenario than that described in
Section~\ref{sec:course-spline} for the elevation profile plot, where
values of $\xi$ come directly from the plot. Instead, we need to compute
the $\xi$ coordinates of each point using a procedure analogous to what
was described for the 3D GPS data. To this end, we first construct
splines $S(\alpha)$ and $Z(\alpha)$ that interpolate the extracted data
points, then use Simpson's rule to approximate the corresponding $\xi$
values from the integral
\begin{gather*}
  \xi_i = \xi(\alpha_i) \approx  \int_0^{\alpha_i} 
  \ds\sqrt{S_\alpha^2 - Z_\alpha^2}\; d\alpha,
\end{gather*}
after which the required Hermite splines for $Z(\xi)$ and $S(\xi)$ can
be built.  The {\tt makima} interpolant for elevation is depicted in
Fig.~\ref{fig:moxnes-sims}a, alongside the linear spline fit used by
MSH.

MSH provide no indication whether their course is FIS-rated, but it is
nonetheless interesting to verify whether it satisfies any of the
homologation criteria in Section~\ref{sec:homologation}. Starting with
criterion H2 for example, the arc length of the Hermite spline
interpolant is 4218~m which sits within 0.5\% of the stated 4.2~km
length. The average gradient is 8.1\% (computed as the mean value of
$|\theta|$), which sits within the allowable range of 6--14\%\ (H3). The
point-wise gradients also lie between $-$23 and $+$24\%, which exceed
the FIS limit of $\pm$18\%, but only on short sections and so this is
also allowed (H4).  Being able to easily test these criteria is a useful
by-product of our interpolation approach.

We are now ready to compare our model with the baseline test case from
MSH.  For a linear spline interpolant, the MSH simulation required a
time of 815~s to complete which they found differed from the
experimentally measured time by 2 seconds (although they did not state
whether the actual time was 813 or 817~s). Applying our \Matlab\ code to
solve the ODE system~\eqref{eq:ode2a}--\eqref{eq:ode2b} with the same
linear interpolant requires 813~s of skiing time, which is within the
range reported by MSH.  They also provided no details about their ODE
solver, so we can expect that some discrepancies will arise from
differences in the numerical approximations being used.  One advantage
of using a robust ODE solver like \Matlab's {\tt ode113} is that
it integrates seamlessly through derivative discontinuities in the ODE
right-hand side arising at spline points.

Although MSH performed most simulations on a piecewise linear course,
they did include a single run with a cubic spline interpolant for which
their skier takes roughly 20~s longer (or 835~s, estimated from
\cite[Fig.~16]{moxnes2014using}).  In comparison, when we run a second
simulation with the Hermite spline interpolant, the skiing time
increases to 823~s. This time lies roughly midway between those for
MSH's linear/cubic results, which is to be expected since the cubic
spline should give the longest arc length (owing to its more oscillatory
nature) and the linear spline the shortest.

This 10--20~s variation in skiing times between the linear and smooth
interpolants is a 1--2.5\%\ difference, which seems relatively small and
indeed, MSH argued that this difference is negligible. The slight
difference in course length can be seen in the elevation plot from
Fig.~\ref{fig:moxnes-sims}a, and the corresponding plot of speed versus
time in Fig.~\ref{fig:moxnes-sims}b shows how $v(t)$ follows the same
trend for both courses. There are some small discrepancies as well as a
noticeable lag in the {\tt makima} result by the end of the race,
although for all simulations the average speed remained consistent at
about 5.1~m/s. An animated view of these results is included as a MPEG
video in the Supplementary Materials.

Nevertheless, to obtain realistic simulations in a race context where a
second or two can distinguish between first and second
place\footnote{Consider this quote from Hinder et
  al.~\cite{hinder-etal-2024}: ``During the 2022 Games in Beijing, a
  1\%\ improvement in the finishing times of 31 athletes who came in
  from second to eighth place would have enabled them to win a gold
  medal instead.''}, we should strive for as much accuracy as possible
in order to differentiate skier dynamics over a range of course
conditions and skiing techniques. Because modifying the spline
interpolant requires a trivial one-line change to the code, we argue
that it is always preferable to employ a smooth Hermite spline
interpolant that provides a more accurate and realistic representation
of an actual course.

Finally, we include a plot of the force components during our Hermite
spline simulation, along with the total force
$\subtext{F}{tot}=F_p+F_g +F_s+F_d$.  Propulsive and gravitational
forces clearly dominate throughout, with the largest peaks in $F_p$
arising on steep uphills and coinciding with lower speeds and large
(negative) values of $F_g$.  Conversely, steep downhills lead to much
larger $v$, positive $F_g$, and lower $F_p$.  The snow friction force
has a smaller magnitude on average, but in contrast with other forces it
induces a relatively steady contribution opposing forward motion over
the entire course. Aerodynamic drag has the smallest impact on skier
dynamics, but does spike over short periods on the steepest
downhills. It is interesting to observe how the total force oscillates
about zero, and that there are regular time intervals where
$\subtext{F}{tot}\approx 0$ corresponding to sections with relatively
constant slope (such as the steep uphill section with $t\in[250,310]$
and $\xi\in[1000,1150]$, depicted in Fig.~\ref{fig:moxnes-sims}d). These
are stretches where the athlete is able to work themselves up to a
roughly constant speed, after which all forces are in balance and hence
acceleration drops to zero.

\begin{exercise}
  \label{ex:sims2D-hat}
  Modify the code {\tt skirun2d.m} to simulate two idealized hill shapes
  that approximate what a skier would normally encounter on an actual
  course.
  \begin{enumerate}[label=(\alph*)]
  \item Consider a piecewise linear hill described by
    $z(\xi) = 40 + \frac{1}{10}\big| \xi-400 \big|$ for
    $0\leqslant \xi \leqslant 800$.  This is analogous to ``gluing
    together'' end-to-end the up/downhill sections from
    Exercise~\ref{ex:exact2D-hat}b--c, with each stretch having length
    $40\sqrt{101} \approx 402$~m.  Implementing this in \Matlab\ is
    easiest using the {\tt interp1} function with option {\tt 'linear'}
    to interpolate the 3 data points $(0,0)$, $(400,40)$ and
    $(800,0)$. Use the same parameters as in
    Exercise~\ref{ex:exact2D-hat} and compare the computed speed and
    completion time with the exact solution.
  \item Next, take a smoothed version of the course from part (a) with
    $z(\xi) = 40\sin\left( \frac{\pi \xi}{800} \right)$, and interpolate
    using 40 equally-spaced spline points on the interval
    $0\leqslant \xi \leqslant 800$.  Before running any simulations,
    consider this question: If a skier races on this hill and the one
    from (a), while applying the same constant propulsion force, which
    course will they complete faster?  Then use the code to simulate a
    skier on the smooth hill to determine whether your intuition was
    correct.
  \end{enumerate}
\end{exercise}

\begin{exercise}
  \label{ex:drag}
  Investigate how important it is to have a speed-dependent switch in
  the drag coefficient between tuck and upright positions. Compare the
  baseline simulation for a skier on the MSH course using $C_dA(v)$ from
  Eq.~\eqref{eq:CdA}, with a second simulation for a constant
  $C_dA\equiv 0.45$, which is just the average of the tuck/upright
  values.
\end{exercise}

\begin{exercise}
  \label{ex:mass+power}
  Consider this quote from Carlsson et al.~\cite{carlsson-chapter-2016}:
  ``Generally a smaller skier is favored on uphills and the larger skier
  is favored on dowhills and the flat.''  Investigate by varying skier
  mass $\pm$15~kg from the baseline value and comparing the finish times
  with those in MSH~\cite[Fig.~11]{moxnes2014using}. Changing mass by
  itself ignores the fact that more muscular (heavier) skiers can
  generate higher peak propulsion force. So, repeat your simulations
  with $\Pmax=275\pm 55$~W, assuming the same two skiers generate
  $\pm$20\% of the baseline power.  This connection between mass and
  power is discussed further by Carlsson et
  al.~\cite{carlsson-etal-2011}.
\end{exercise}

\begin{exercise}
  \label{ex:ODEsolvers}
  Compare the accuracy and computational cost when \Matlab's ODE solvers
  {\tt ode113}, {\tt ode45}, {\tt ode15s} and {\tt ode23s} are applied
  to the ``baseline'' MSH example. Use the difference in race times as a
  simple measure of accuracy, and compare computational cost using the
  built-in functions {\tt tic}, {\tt toc}.  Discuss the results, paying
  particular attention to the algorithms implemented in the four
  solvers as described in the \Matlab\ documentation.
\end{exercise}

%%%%%%%%%%%%%%%%%%%%%%%%%%%%%%%%%%%%%%%%%%%%%%%%%%%%%%%%%%%%%%%%%%%%%%
\subsection{Example 2: Welde's Measurements of Elite Racers}
\label{sec:2Dsims-Welde}

As a second example, we consider an experimental study by Welde et
al.~\cite{welde2017pacing} that reports results for 36 Norwegian skiers
racing on a 15~km course in Troms\o, Norway, using the classic
technique.  Their study (which we'll refer to as the ``Welde
experiment'') involved skiers with a range of abilities who they
classified into two equal-sized groups of ``slow'' and ``fast'' skiers.
The fast group included several top-10 World Cup finishers, and so this
example affords us with an excellent opportunity to further validate our
model with high-performance skiers and determine whether the power
function (specifically $\Pmax$) is able to differentiate elite skiers
from other competitors.

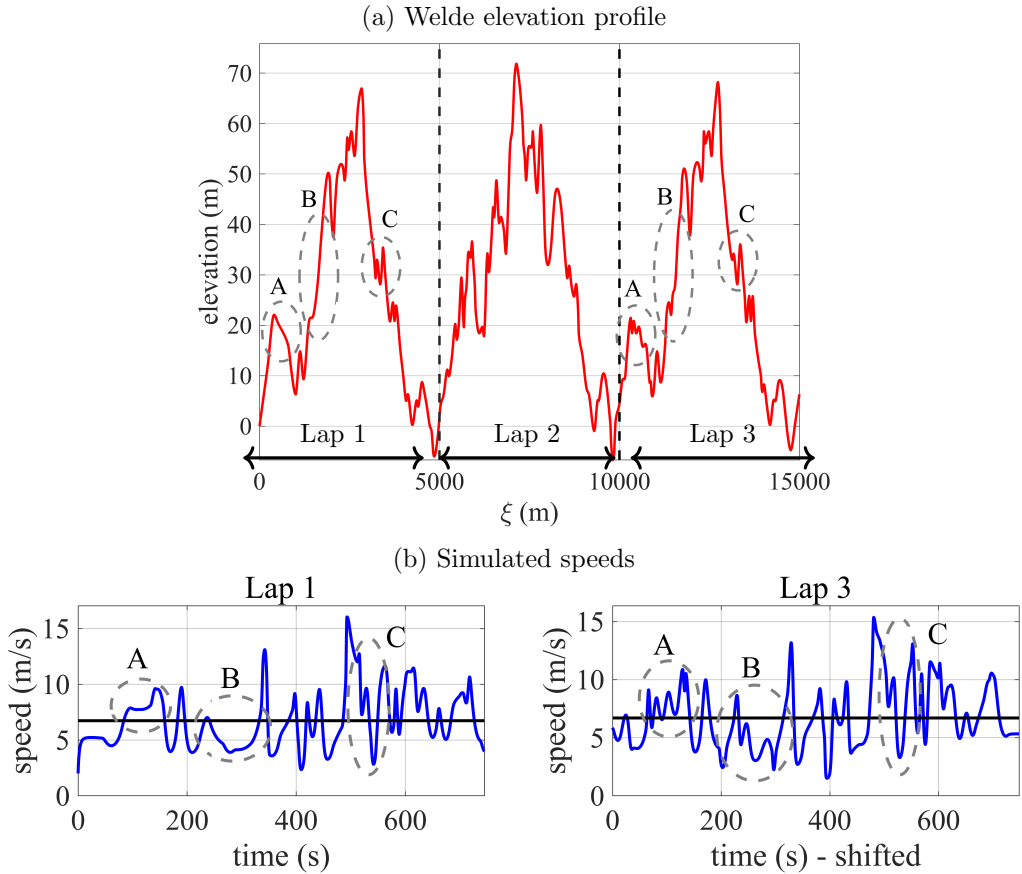
\begin{figure}[hbt]
  \centering
  (a) Welde elevation profile\\
  \begin{tikzpicture}[scale=1]
    \node at (0,0) {\includegraphics[width=0.5\textwidth]{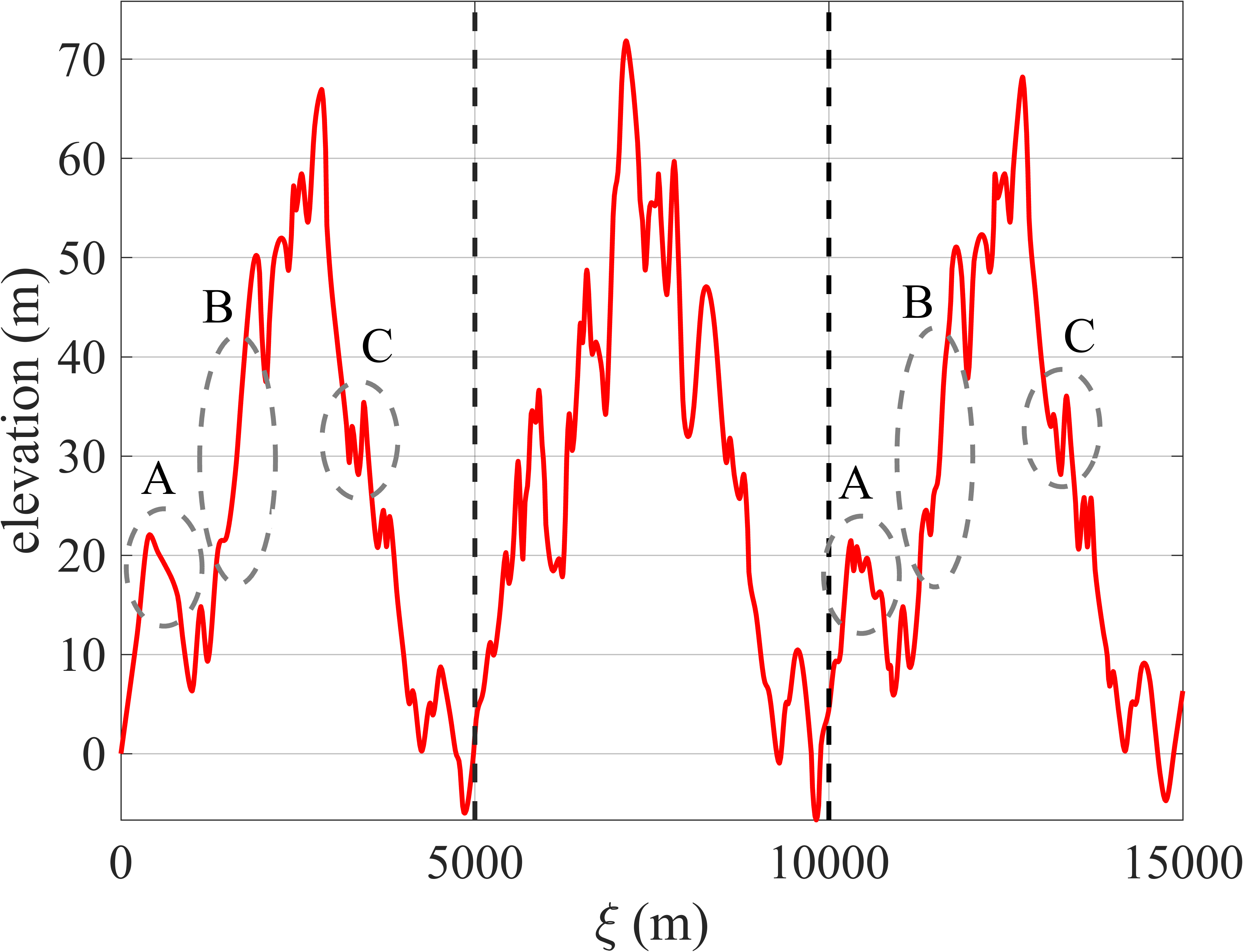}};
    % \draw[black,very thick,dashed,opacity=0.7] (-1.1,-2.5)--(-1.1,3.4);
    % \draw[black,very thick,dashed,opacity=0.7] (1.4,-2.5)--(1.4,3.4);
    \draw[black,very thick,<->] (-3.6,-2.3) -- (-1.2,-2.3) node[midway,above] {Lap 1};
    \draw[black,very thick,<->] (-1.0,-2.3) -- (1.35,-2.3) node[midway,above] {Lap 2};
    \draw[black,very thick,<->] (1.55,-2.3) -- (4.0,-2.3) node[midway,above] {Lap 3};
  \end{tikzpicture}
  \\
  (b) Simulated speeds\\
  \includegraphics[width=0.8\textwidth]{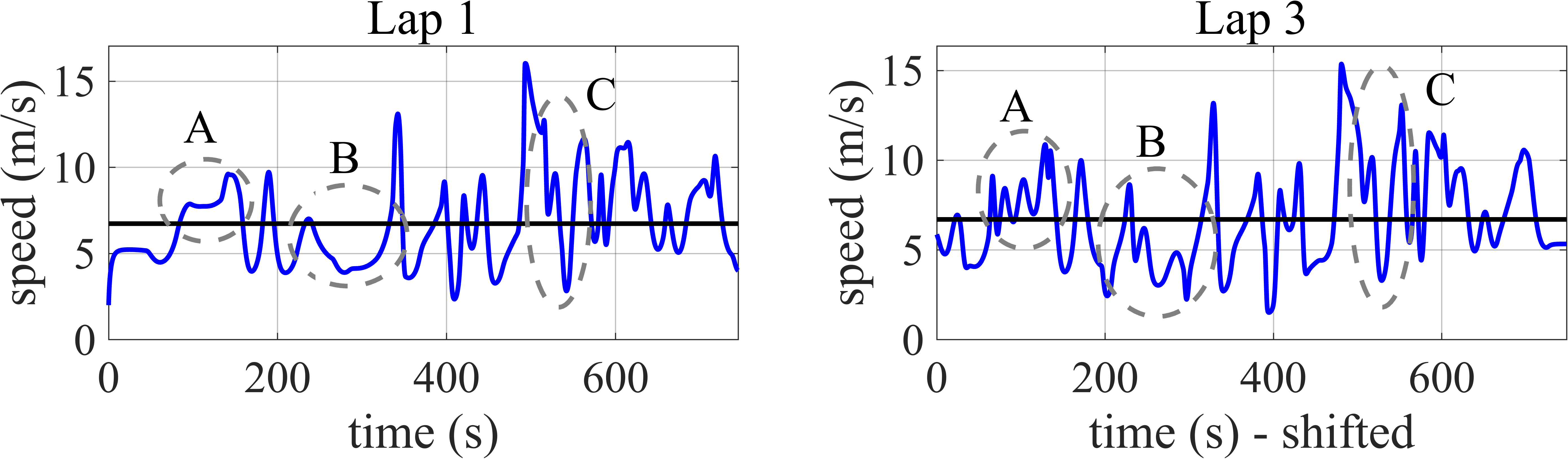}
  \caption{Simulations of the Welde course, taking all parameters equal
    to the baseline values except $\Pmax=434$, corresponding to the time
    for the winner of the race. (a, Top) Elevation profile interpolated
    with the Hermite spline. Numbered circles indicate clear
    discrepancies between laps 1 and 3, which are identical on the
    actual course. (b, Bottom) Simulations of the skier's speed during
    laps~1 and~3 (with the time for lap~3 shifted so it also starts at
    $t=0$). The black horizontal lines at $v=6.7$~m/s denote the average
    lap speed.}
  \label{fig:weldeSims}
\end{figure}

The authors supply a 2D elevation plot~\cite[Fig.~2]{welde2017pacing}
for their course, which is divided into three laps of length 5~km each,
with laps 1 and 3 being repeated runs of the same loop.  The $(\xi, z)$
coordinates can be extracted directly from their course plot using
WebPlotDigitizer, in contrast with the $(s,z)$ plot from Example~1.
Because this course is so much longer than the previous example, we
decided to exploit the ``automatic extraction'' feature of
WebPlotDigitizer rather than manually selecting the points.  This
generates a CSV file with 203 data points that are used to build the
Hermite spline interpolant pictured in Fig.~\ref{fig:weldeSims}a.  It is
important to note that the Welde elevation plot is under-resolved
relative to that from Example~1, not only since the course is over three
times longer, but also because the $z$-axis scale in the plot image is
so much smaller.  The reduced resolution introduces several anomalies in
the spline interpolant that are easily identified by comparing laps 1
and 3 in Fig.~\ref{fig:weldeSims}a, and focusing on the sections
labelled A,B,C.

Welde's experiment was based on a high-level race during the Norwegian
national championships, but the authors did not indicate whether the
course was FIS-certified; nonetheless, we will still discuss whether the
homologation requirements are met.  Criteria H2 and H3 are easily
satisfied since the arc length of the spline interpolant is 15048~m
(within 0.3\% of the stated length) and the average gradient is 8.4\%
(similar to MSH). However, the gradient varies between $-$69 and $+$47\%
over the course, based on a fine-grid sampling of the spline
$\Theta(\xi)$, which indicates some steep up- and downhills that might
not pass FIS certification. On closer inspection, none of these gradient
anomalies occur at or near spline data points but appear instead as
spikes or oscillations near the center of a spline interval.  This
suggests that these spikes are an artifact of the spline interpolation,
which likely arise from data extraction errors, or differentiating the
spline, or both. But despite certain fine-scale course features that may
not be accurately captured, they are few in number and restricted to
relatively small sections, and so we are still confident that the
Hermite spline provides a reasonable representation of the actual
course.

\begin{table}[b!]
\centering
\caption{Summary of Welde's data from 36 racers in a 15~km classic ski
  race in Troms\o, Norway~\cite[Supporting data file
  S1]{welde2017pacing}.  The rows list skiers divided into two groups of
  ``slow'' and ``fast'' skiers numbering 18 each, with the winner's
  result listed separately. The columns list race time (in s) and
  average speed (in m/s) for the full course, with speeds for laps~1
  and~3 included separately.  The final column indicates the power
  parameter $\Pmax$ needed in order that our model simulations match the
  time and speed to ski the full course.}
\label{tab:welde}
{\renewcommand{\arraystretch}{1}
  \begin{tabular}{l|cc|cc|l}
    Skier & \multicolumn{2}{c|}{Full course} & Lap 1 & Lap 3 & Model\\
    class & time & speed & speed & speed & $\Pmax$ \\\hline
    Slow  & 2560 & 5.86 & 6.22 & 5.83 & 346 \\
    Fast  & 2389 & 6.30 & 6.62 & 6.28 & 392 \\
    Winner& 2268 & 6.61 & 6.95 & 6.60 & 434 
  \end{tabular}}
\end{table}
We now describe some specific race results reported by Welde and
summarized in Table~\ref{tab:welde}. They observed that the bottom 18
``slow'' skiers completed the race in an average time of 2560~s (with
average speed 5.86~m/s) whereas the top 18 ``fast'' skiers finished in
2381~s (average speed 6.30~m/s).  The difference in performance between
fast and slow skiers is largely a function of their fitness level,
technique used, and more specifically the power levels they can sustain
throughout a race. We can incorporate this fast/slow difference in our
model by manually adjusting the parameter $\Pmax$ in the power
function~\eqref{eq:Pv} until a simulation yields the desired race time.
Using this approach, we find that choosing $\Pmax=392$ matches the
average time for fast skiers (to within 1~s), whereas the slow time is
obtained with $\Pmax=346$. Welde also singles out the race winner who
finished well ahead of all other competitors with a time of 2268~s, that
is captured (to within 1~s) by our model by taking $\Pmax=434$.  This
clearly identifies the winner as an outlier, since the difference in
maximum power between the winner and average fast skier is almost as
large as the difference between the fast and slow groups.  This large
discrepancy is most likely attributable to the winner's exceptional
technique, based on Welde's observation that the winner was the only
skier who used double-poling (a very energy- intensive technique)
throughout the entire race.  We also note that the elite skier from
Example~1 had a much lower value of $\Pmax$ than all three groups in
Welde's study, which suggests that the self-chosen pace of the MSH
athlete was significantly slower than their race pace.

The simulations for the winner are depicted in Fig.~\ref{fig:weldeSims}b
in terms of the skier speed during laps 1 and 3, which exhibit
noticeable variations especially in the three highlighted sections
labelled A,B,C. Since these are identical course sections, these
differences are most likely due to variations in the spline interpolants
for laps 1 and 3. The average speed is indicated by a horizontal line in
Fig.~\ref{fig:weldeSims}b, which is essentially identical at 6.7~m/s for
both laps. This should be contrasted with the Welde results in
Table~\ref{tab:welde}, where the winner's average speed drops by a wide
margin of 5\% between the first and last laps. The discrepancies
identified in the speed plots for the two laps are too small and short-
lasting to cause such a large reduction in speed.  This slowdown between
laps 1 and 3 is described by Welde as ``positive pacing'' and can be
thought of as a consequence of muscle fatigue, in which the athlete
slows down throughout a race as they exhaust their energy reserves and
muscles tire. As currently formulated, our model with constant $\Pmax$
is not capable of capturing this fatigue effect; however, we could
easily emulate racer fatigue in a na\"ive way by allowing the peak power
to decrease with time throughout a simulation.  This simple approach to
modelling fatigue is explored in Exercise~\ref{ex:fatigue}, although
fully capturing Welde's results would require a much deeper study that
incorporates experimental observations on positive pacing (e.g.,
\cite{losnegard-kjeldsen-skattebo-2016,
  stoggl-pellegrini-holmberg-2018}).

% Comparing our results to the experimental output, one shortcoming of
% our model is that on this course where there is a repeat of the first
% 5000 meters at the end of the race, our velocity curve measures in
% similar magnitudes along both laps, and any variation across these two
% curves comes from the inaccuracy from our data collection method. The
% times to finish the 5000 meters in the first and last lap, are
% approximately 746 s and 748 s, respectively, measuring only 0.3\%
% slower the second time around (Fig.~\ref{fig:weldeSims}). This is a
% stark contrast to the the recorded velocity of skiers as 11.8\% slower
% on the final lap due to the ``positive pacing" strategy employed by
% skiers. We elaborate on this shortcoming in
% Section~\ref{sec:discussion}, where we discuss future projects and
% incorporating fatigue into the model.

% \begin{exercise}
%   \label{ex:welde-time}
%   The `fast' skiers in Welde's experiment weighed between $[74.4, 80.8]$ kg. 
%   Adjust the power parameter $\Pmax$ for weights at the lower and higher end of 
%   this interval until the race times match the winner's race time within an 
%   accuracy of 1~s. 
% \end{exercise}

\begin{exercise}
  \label{ex:fatigue}
  Notice from Table~\ref{tab:welde} that skiers are 5--6\% slower on
  lap~3 compared to lap~1, which suggests they may be experiencing
  fatigue owing to their muscles tiring throughout the course. A simple
  way to incorporate this effect is to assume that halfway through the
  race (for $\xi\geqslant 7500$~m), maximum power drops from $\Pmax$ to
  $c\Pmax$, where the constant $0<c<1$ can be thought of as a ``fatigue
  factor''.  Modify {\tt skirun2d.m} to incorporate muscle fatigue in
  this manner.  Then adjust the values of $c$ and $\Pmax$ until the
  total race time for the ``winner'' is the same as in
  Table~\ref{tab:welde}, and their speed on lap~3 is 5--6\% slower than
  on lap~1.
\end{exercise}

% \begin{exercise}[More of a project] 
%   Investigate generating a ``smoother'' spline that 
%   eliminates the anomalously large slopes in the Welde course. 
%   Try using a different spline interpolant for the course (or specifically for 
%   $\Theta(\xi)$, or adjusting the data points (perhaps with a mild smoothing 
%   algorithm), or some other approach.
% \end{exercise}

%%%%%%%%%%%%%%%%%%%%%%%%%%%%%%%%%%%%%%%%%%%%%%%%%%%%%%%%%%%%%%%%%%%%%% 
\subsection{Example 3: Simulating the Toblach Ole Course in 2D}
\label{sec:2Dsims-Ole}

One of the strengths of our modelling approach is that it can easily
handle courses specified as 2D elevation plots or 3D GPS data. Back in
Section~\ref{sec:course-spline}, we introduced the 4.2~km Ole course in
Toblach, imported from a GPS file\footnote{Toblach (or Dobbiaco in
  Italian) is located in the South Tyrol region of the Italian Alps (or
  Dolomites), and lies within a UNESCO World Heritage Site containing 18
  major peaks, that also contains the winter resort town of Cortina
  d'Ampezzo.  Cortina hosted the Winter Olympic Games in 1956, and will
  do so again for the second time in 2026.}. This is one of several
FIS-rated courses for which a file of GPS waypoints is posted on the
Dolomiti NordicSki web site~\cite{dolomiti-gps-2023}.  We explained how
to convert GPS data $(\latitude, \longitude, z)$ to Cartesian
coordinates, which can then be approximated by spline interpolants
$X(\xi)$, $Y(\xi)$, $Z(\xi)$ that are parameterized by projected
distance $\xi$. The Ole course provides an ideal opportunity to test our
2D model on a course derived from real three-dimensional data, and then
after extending the model to 3D later in Section~\ref{sec:3Dmodel}, we
can compare with simulations that take into account inherently 3D
effects such as braking on steep downhills.

After computing the arc length and inclination angle (using the code
{\tt setup3d.m}), it is easy to verify that the Ole course satisfies the
homologation criteria H2--H4. The length of the spline is 4157~m and the
average gradient is 7.9\%, both of which are well within the stated
limits (for H2 and H3). The pointwise gradients also remain between
$\pm$18\% as required by H4, except for a few short sections where it
spikes to 40\%. Regarding criterion H5, the last 100~m of the course are
almost completely flat which is not ideal but may still be considered
acceptable; however, the start begins immediately with a 5${}^\circ$
uphill slope that clearly violates H5 (this issue was addressed
previously in Exercise~\ref{ex:start-missing}).

\begin{figure}[tb]
  \centering
  \includegraphics[width=0.6\textwidth,trim=0 0 10 0,clip]{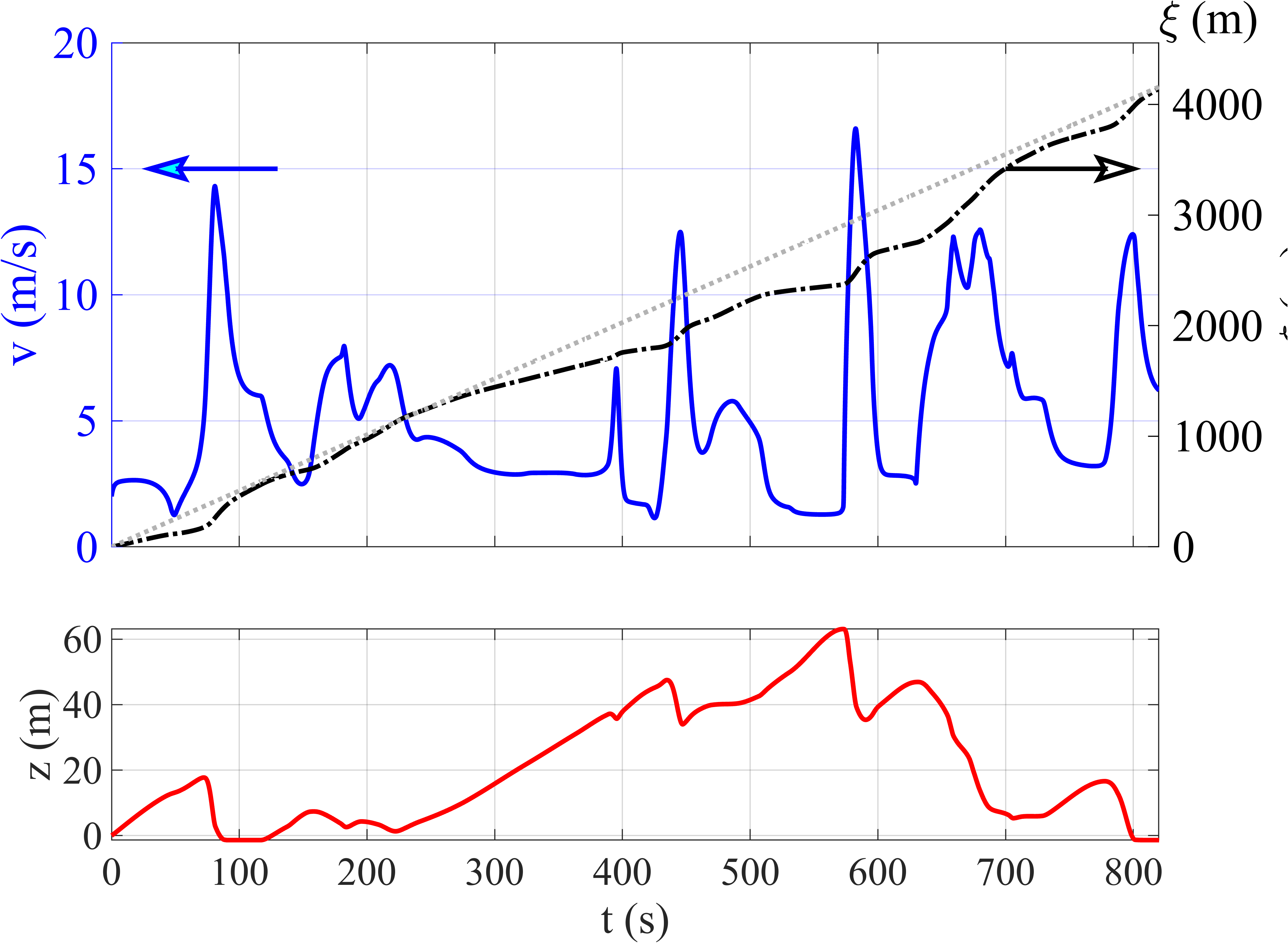}
  \caption{Simulation of the 4.2~km Ole course with the baseline
    parameters, displaying speed and projected arc length (top plot,
    blue/black lines) and elevation (bottom plot, red line). The grey
    dotted line in the top plot is $\xi(t)=5.07\, t$, corresponding to a
    skier moving at the constant average speed.} \label{fig:ole2D}
\end{figure}

Simulating the Ole course using the baseline parameters leads to a
skiing time of 819.9~s corresponding to an average speed of 5.07~m/s,
which is very close to what we computed on the 2D MSH course in
Example~1.  This is not surprising because, for two courses based on
similar design criteria, the skier's speed should be largely a function
of power output, and $\Pmax$ is the same in both examples.
Fig.~\ref{fig:ole2D} depicts the elevation, speed and projected arc
length as functions of time $t$.  This view of the solution makes it
easy to connect steep up/downhill sections with intervals where speed is
lower/higher. Also, the progress of skier distance $\xi$ is fairly
consistent and doesn't deviate very far from the straight line with
slope equal to the average speed, but the most rapid changes in $\xi$
are easily connected with steep downhills (and slowest changes with
uphills).

\begin{exercise}
  \label{ex:albertSprint}
  On the Dolomiti NordicSki website~\cite{dolomiti-gps-2023}, find the
  elevation profile plot for the 1.4~km ``Albert Sprint'' course (also
  called ``Stadium Track''), which is a sequence of two hills having
  roughly similar size and shape. Use
  WebPlotDigitizer~\cite{webPlotDigitizer} to extract a sample of
  $(\xi,z)$ points with roughly 100~m spacing, write them to a CSV file,
  and use them to simulate a race with the baseline parameters. Next,
  consider a second manufactured course built from a sum of two
  ``Gaussian bump'' functions of the form
  $a \exp \left(-\frac{1}{c}(\xi - b)^2\right)$, and tune the parameters
  $a,b,c$ to most closely match the first course. Generate CSV data with
  similar resolution and repeat the simulation.  How do the two results
  compare?
\end{exercise}

\begin{exercise}
  \label{ex:carlsson}
  Carlsson et al.~\cite{carlsson-etal-2011} developed a model very
  similar to ours, and used it to simulate skiing over a simple hill
  that consists of a moderate uphill slope (with constant inclination
  angle $\theta=+3^\circ$) followed by a steep downhill (with constant
  $\theta=-20^\circ$), and a smooth transition between the two. Extract
  data from the elevation plot in~\cite[Fig.~3]{carlsson-etal-2011},
  making sure that the gradient at both start/finish is zero. Then
  simulate a skier using these parameters: skier mass $m=80$~kg, snow
  friction $\mu=0.03$, and maximum power $\Pmax=366.5$~W. Plot the
  solution and show that the skier completes the course in roughly
  136~s.  Then compare with the time required to run the course in
  reverse (right-to-left), starting instead on the steep uphill
  section. Is this consistent with your intuition?
\end{exercise}

\begin{exercise}
  \label{ex:wind}
  Investigate skiing in windy conditions, and assume there is a constant
  head/tailwind with speed $\vwind$.  Modify the aerodynamic drag force
  appropriately, and then repeat the simulations with the $\vwind$
  values displayed in MSH's Figure~14.  When skiing against a headwind,
  at which value of $\vwind$ does the aerodynamic drag force start to
  dominate over other forces?
\end{exercise}

%%%%%%%%%%%%%%%%%%%%%%%%%%%%%%%%%%%%%%%%%%%%%%%%%%%%%%%%%%%%%%%%%%%%%%
% ---------------------------------------------------------------------
%%%%%%%%%%%%%%%%%%%%%%%%%%%%%%%%%%%%%%%%%%%%%%%%%%%%%%%%%%%%%%%%%%%%%% 
\section{Skiing in Three Dimensions}
\label{sec:3Dmodel}

\subsection{Curvature and Braking Force}
\label{sec:3Dbraking}
\begin{figure}[tb]
  \centering  %\includegraphics[width=0.5\textwidth]{skigeometry3d}
  \begin{tikzpicture}[scale=0.55]
    \draw[gray!20!white,fill] (0,0) -- (15,0) -- (11,-3.2) -- (-5.6,-3.2) -- (0,0);
    \draw[gray!20!white,fill] (0,0) -- (15,0) -- (16,1) -- (1.7321,1) -- (0,0);
    \node[above right,inner sep=0pt,scale=1.3,red!70!black,] at (10,2.3) {$\vec{r}(\xi)$};
    % Vertical dotted lines
    \draw[dotted,thick] (6.2,4.252) -- (6.2,-1.063);
    \draw[dotted,thick] (10.95,0.8112) -- (10.95,-2.993);
    \draw[dotted,thick] (8.2,-2.7) -- (8.2,5.952);
    \draw[dotted,thick] (1.55,0.65) -- (1.55,1.9); %%%%%
    \draw[dotted,thick] (3.622,-0.25) -- (3.622,1.343);
    % Projected path
    \draw[magenta!70!pink,very thick,dash pattern=on 3pt off 2pt] 
    (0,0) to[spline through={ (1.7, -0.4) (3.622, -0.2) (3.1, -0.15)
      (2,0) (1.538, 0.5) (1.7, 0.8) (2.3, 0.8) (3, 0.6) (4.5,0.3)
      (5.5, -0.2) (6.200, -0.9) (6.922, -1.427) (8.263, -1.063)
      (9.226, -1.063) (9.653, -1.371) (10.13, -1.622) (9.918, -1.762)
      (9.461, -2.070) (9.671, -2.601) (10.29, -2.797) }] (10.95, -2.993); 
    % Skier path
    \draw[red!70!black, ultra thick] (0,0) to[spline through={
      (2.144, 0.3636) (3.405, 0.9231) (3.622, 1.343) (3.167, 1.287)
      (1.875, 1.307) (1.578, 1.992) (1.811, 2.485) (2.563, 2.777) (4.194, 2.853)
      (5.482, 3.301) (6.200, 4.252) (6.909, 4.587) (7.711, 4.420)
      (8.481, 3.524) (9.574, 2.909) (9.917, 2.182) (9.513, 1.762)
      (8.949, 1.427) (9.241, 1.175) (10.20, 0.9510) }](10.95, 0.8112); 
    % Ellipse for radius of curvature
    \draw[mygreen,fill=mygreen!10!white,dash pattern=on 2pt off 1pt,thick,rotate=18] (3.0,1.15) ellipse (0.95 and 0.7);
    \draw[thick,mygreen,-{Straight Barb[length=2mm,width=1.5mm]}]
    (2.45,2) -- (1.55,2.1) node[very near start,above,inner
    sep=3pt,scale=0.8] {$R$}; 
    \draw[mygreen,fill,very thick] (2.5,2) circle (0.8mm);
    \draw[mygreen,fill=white,very thick] (1.55,2.1) circle (1.5mm);
    \draw[magenta!70!pink,fill,very thick] (1.55,0.65) circle (0.8mm);
    % Axes
    \draw[thick,-{Stealth[length=4mm]}] (0.3,0.175)--(-6,-3.5) node[left,scale=1.3,inner sep=1pt] {$x$};    
    \draw[thick,-{Stealth[length=4mm]}] (-0.35,0)--(16,0) node[right,scale=1.3,inner sep=1pt] {$y$};
    \draw[thick,-{Stealth[length=4mm]}] (0,-0.35)--(0,4) node[above,scale=1.3,inner sep=1pt] {$z$};
    % Arrows, lines and labels
    \draw[blue,very thick,dashed] (6.2,4.252)--(6.2,6.0);
    \draw[blue,very thick,dashed] (6.2,4.252)--(7.5,3.2);
    \draw[blue,-{Stealth[length=4mm]},very thick] (6.2,4.252)--(8.2,5.952) node[above right,scale=1.3,blue,inner sep=0pt] {$v$};
    \draw[blue,->] (7,3.6) arc (-30:30:1.3) node[right,scale=1.3,inner sep=0pt] at (7.2,3.9) {$\inclination$};
    \draw[blue,->] (6.2,5.3) arc (90:38:1) node[right,scale=1.3,inner sep=0pt] at (6.5,5.8) {$\widetilde{\inclination}$};
    %
    %\draw[black,thick,dotted,-{Stealth}] (6.2,-0.98) -- (4.9,-2.063) node[left,inner sep=1pt] {$x$};
    \draw[blue,very thick,dashed] (6.2,-0.98) -- (4.9,-2.063);
    \draw[blue,-{Stealth[length=2.5mm]},thick] (6.2,-0.98)--(8.2,-2.8);
    \draw[blue,->] (5.5,-1.5) arc (-120:-70:1.5)
    node[right,scale=1.3,inner sep=0pt] at (5.9,-2.3) {$\azimuth$}; 
    % Points
    \draw[red!70!black,fill] (0,0) circle (1mm);
    \draw[red!70!black,fill] (10.95, 0.8112) circle (1mm);
    \draw[magenta!70!pink,fill] (10.95, -2.993) circle (1mm);
    \draw[blue,fill=white,very thick]  (6.2,4.252) circle (1.5mm);
    \draw[blue,fill=white,very thick]  (6.2,-0.98) circle (1.5mm);
    \draw[magenta,thick,dotted,-{Straight Barb[width=1.5mm]}] (0.4,-0.35) arc (-115:-98:4) node[below,scale=1,inner sep=1pt] at (1,-0.6) {$\xi$}; 
    \draw[red!70!black,thick,dotted,-{Straight Barb[width=1.5mm]}] (0.2,0.3) arc (-90:-75:4.3) node[above,scale=1,inner sep=1pt] at (0.8,0.4) {$s$}; 
  \end{tikzpicture}
  \caption{Definition of the inclination angle $\inclination$ (measured
    relative to the horizontal plane) and azimuth angle $\azimuth$ (in
    the $x,y$-plane, measured relative to the $x$-axis) for a 3D ski
    course parameterized as
    $\vec{r}(\xi) = \big(x(\xi),\, y(\xi),\, z(\xi)\big)$.  The radius
    of curvature $R=\frac{1}{\kappa}$ is also shown as the radius of the
    circle (dashed, green) that best approximates the curve at any given
    point.}
  \label{fig:theta-phi-3D}
\end{figure}
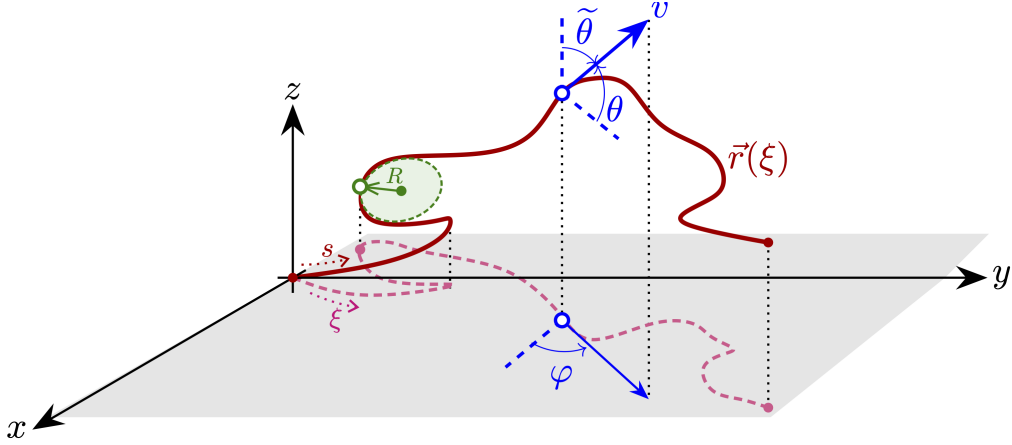

We now extend the 2D model to a more realistic setting that captures the
dynamics of an athlete skiing along a 3D course, who is turning from
left to right on top of dealing with changes in elevation on hilly
sections. We have already explained in Section~\ref{sec:course-spline}
the algorithms behind constructing the Hermite spline approximation of a
3D course. Recall that such a course can be written as a vector function
$\vec{r}(\alpha) = \big(x(\alpha),\, y(\alpha),\, z(\alpha)\big)$, where
we will employ two choices for the parameter: $\alpha=\xi$ when
referring to the course geometry, and $\alpha=t$ when formulating the
ODEs governing skier dynamics. The diagram in
Fig.~\ref{fig:theta-phi-3D} depicts an idealized course $\vec{r}(\xi)$
along with its projected path in the $x,y$-plane, and highlights an
arbitrary point where the skier is moving with speed $v$ in the tangent
direction.  The inclination angle $\inclination$ is measured relative to
the horizontal as in 2D, but must be viewed within the vertical plane
containing the tangent vector. In 3D, we require an additional azimuth
angle $\azimuth$ that captures the horizontal skiing direction measured
relative to the $x$-axis.  Note that $\azimuth$ has the same definition
as in 3D spherical coordinates, whereas the inclination angle is related
to the usual spherical polar angle $\widetilde{\inclination}$ by
$\inclination = \frac{\pi}{2}- \widetilde{\inclination}$. Using standard
formulas from vector calculus, the angles may be expressed in terms of
the coordinate functions as\footnote{A detailed discussion of spherical
  angles can be found in the
  \href{https://en.wikipedia.org/wiki/Spherical_coordinate_system}{Wikipedia
    article ``Spherical coordinate system''}.}
\begin{gather}
  \inclination = \arctan \Bigg( 
  \raisebox{0.2cm}{$\ds\frac{z_\xi}{\sqrt{x_{\xi}^2 + y_{\xi}^2}}$} \Bigg) 
  \quad\text{and}\quad
  \azimuth = \operatorname{sign}(y_\xi) \arccos \Bigg(
  \raisebox{0.2cm}{$\ds\frac{x_\xi}{\sqrt{x_\xi^2 + y_\xi^2}}$} \Bigg).
  \label{eq:az-el}
\end{gather}
When building our spline approximations in 3D, we proceed as in 2D and
use a fine $\xi_i$ grid to compute values for both $\inclination_i$ and
$\varphi_i$, and then construct the corresponding Hermite interpolants
$\Theta(\xi)$ and $\Phi(\xi)$. These splines can be precomputed and then
cheaply evaluated to obtain the two direction angles during a simulation
using the current value of $\xi(t)$.

A major difference between skiing in 2D and 3D is that the athlete has
the freedom to turn left or right in 3D, thereby introducing an extra
centripetal force $ma_c$ that acts radially outward (and perpendicular
to the tangent direction). This force is proportional to centripetal
acceleration $a_c={v^2}/{R}$, where $R$ is the radius of curvature at
any location along the course; because $R$ is related to curvature
$\kappa$ by $R = \frac{1}{\kappa}$, an alternate formula is
$a_c=\kappa v^2$.  To counteract the centripetal force and avoid sliding
sideways, the skier must apply an equal and opposite force to their skis
directed radially inward.  In all but the most curvy downhill sections
of the course, this force induced by $a_c$ is very small and easily
countered through minor (mostly unconscious) adjustments in technique.
However, on the steepest and tightest curves (with large $v$ and small
$R$), centripetal effects can become large enough that the athlete can
no longer prevent slipping and risks losing control.  How a skier
handles such turns can be of critical importance in a race as described
by Sandbakk et al.~\cite{sandbakk2014velocity}: ``During the 15-km
pursuit race in the 2010 Olympics in Vancouver, Marit Bj{\o}rgen passed
Justina Kowalczyk in the last downhill turn to win the gold medal.''

In steep downhill turns, skiers reduce their speed by applying a braking
force using one of two primary techniques called the skid turn and step
turn\footnote{In a skid turn, skis are held parallel at a slight angle
  to the direction of motion, and ski edges are scraped against the snow
  to both decelerate and turn.  In a step turn, a skating motion is
  combined with rapid side-to-side stepping to maintain speed as much as
  possible during the turn and dissipate less energy.  There is a third
  technique called the snowplow, which is an extreme version of the skid
  turn and is more of a beginner's strategy that is seldom applied by
  experienced racers.  An in-depth discussion of turning techniques can
  be found in the
  \href{https://nordicskilab.com/nordic-downhill-ski-techniques-for-safety-and-speed}{Nordic
    Ski Lab article ``Nordic downhill ski techniques for safety and
    speed''}.}. In both cases, the braking turn generates a force
opposing forward motion, which we assume following Sandbakk et
al.~\cite{sandbakk2014velocity} is proportional to centripetal
acceleration so that $F_b = \gamma a_c = \gamma \kappa
v^2$, %\label{eq:Fb}
where $\gamma$ is a dimensionless braking coefficient\footnote{A similar
  assumption that braking force is proportional to centripetal
  acceleration is applied to a dynamic model for road cycling
  in~\cite{nee-herterich-2022}.}.
%(see Exercise~\ref{ex:gamma}). 
It is unrealistic for an athlete to apply a braking force throughout an
entire race, even if the force is mostly very small.  Therefore, $F_b$
should be ``thresholded'' so that it is zero except within those
downhill sections where centripetal acceleration exceeds some braking
threshold $\acmin$:
\begin{gather}
  F_b = \begin{cases}
    \gamma \kappa v^2, & 
    \text{if $\theta<0$ and $\kappa v^2 > \acmin$,}\\
    0, & \text{otherwise.}
  \end{cases}
  \label{eq:Fb-threshold}
\end{gather}
The specific choice for parameters $\gamma$ and $\acmin$ will be
discussed separately in Section~\ref{sec:braking}.

Now, the only missing detail is a formula for curvature which is a
standard identity from vector calculus:\footnote{An alternate formula
  that parameterizes curvature as a function of time is obtained by
  replacing $\xi$ with $t$ in Eq.~\eqref{eq:kappa} and substituting
  $\vec{r}^{\,\prime}(t) = \vec{v}(t)$, to get
  $\kappa(t) = \| \vec{v} \times \vec{v}^{\,\prime} \| \big/ v^3$.
  However, it is Eq.~\eqref{eq:kappa} that we implement in our numerical
  simulations.}
\begin{gather}
  \kappa(\xi) = \frac{ \displaystyle \left\| \vec{r}_\xi
      \times \vec{r}_{\xi\xi} \right\|}{
    \displaystyle \left\| \vec{r}_\xi \right\|^3}.
  \label{eq:kappa}
\end{gather}
The derivatives $\vec{r}_\xi$ and $\vec{r}_{\xi\xi}$ appearing
in~\eqref{eq:az-el} and~\eqref{eq:kappa} are easily computed as
derivatives of the Hermite splines $X(\xi)$, $Y(\xi)$ and $Z(\xi)$ using
\Matlab's {\tt fnder} function.  A Hermite spline interpolant is then
built for $\kappa=K(\xi)$, similar to what was done for the elevation
and azimuth angles.

We can now generalize the 2D model
equations~\eqref{eq:ode2a}--\eqref{eq:ode2b} to 3D by adding the braking
force to the ODE for $v$ and multiplying the $\xi$ equation by the
corresponding angular factor from the spherical coordinate
transformation (remembering also to replace
$\sin\widetilde{\inclination} = \sin(\frac{\pi}{2}-\inclination) =
\cos\inclination$). Then the 3D model equations are
\begin{align}
  v^\prime   &= \frac{P(v)}{mv} - g\sin\inclination - \mu g \cos\inclination - 
               \beta v^2 - F_b(v), \label{eq:ode3a}\\
  \xi^\prime &= v\cos\inclination, \label{eq:ode3b}\\
  \intertext{where spline evaluations are used to find
  $\theta$, $x$, $y$, $z$ and $s$ at the current value of $\xi$.  Similar to 2D,
  we can formulate an augmented version of this system that evolves  
  position variables with four additional ODEs:}
%% 
% ~~ \mbox{\footnotesize$\left[~= \sqrt{(x')^2 + (y')^2}~\right]$} 
  x^\prime   &= v\cos\inclination\cos\azimuth, \label{eq:ode3c}\\
  y^\prime   &= v\cos\inclination\sin\azimuth, \label{eq:ode3d}\\
  z^\prime   &= v\sin\inclination, \label{eq:ode3e}\\
  s^\prime   &= v, \label{eq:ode3f}
  % ~~~~~~~~\, \mbox{\footnotesize$\left[~= \sqrt{(\xi')^2 + (z')^2}~\right]$}
\end{align}
where the azimuth angle is obtained by evaluating the spline
$\Phi(\xi)$. In practice, we prefer simulations based on the simpler
two-ODE model \eqref{eq:ode3a}--\eqref{eq:ode3b} because it is more
efficient and also constrains the skier to move exactly along the
parameterized curve describing the course. Both the two- and six-ODE
formulations are implemented in the code {\tt skirun3d.m}, and although
we do not advocate using the augmented model, we still include it for
completeness since it has been implemented by others such
as~\cite{ni-liu-zhang-ke-2022}.

Before moving on, we should mention that some Nordic ski
models~\cite{carlsson-etal-2011, ni-liu-zhang-ke-2022,
  sundstrom2013numerical} incorporate an extra contribution to snow
friction arising from the component of the centripetal force acting
within the $\xi,z$-plane along the direction of travel\footnote{Vector
  calculus tell us that the (signed) curvature for a 2D space curve
  $z(\xi)$ is given by
  $\kappa_{_{2D}}(\xi) = z''(\xi)/(1+z'(\xi))^{3/2}$. Note that this
  formula is equivalent to Eq.~\eqref{eq:kappa}, except that it is a
  signed quantity that is positive/negative depending on whether the 2D
  ' curve is concave up/down.}, which is always directed normal to the
snow surface.  MSH observed that they ``find the greatest differences
between the simulation and the experimental data on terrain with rapidly
changing curvature''~\cite{moxnes2014using}, which we also see at
isolated instants in time. However, because most ski courses have a
balance between up/down-hill sections, the $+/-$ contributions from this
curvature term tend to cancel out on average, leading to a negligible
net effect on the overall skiing time.  Therefore, we have chosen to
ignore the effects of this vertical curvature on snow friction (which is
also consistent with the MSH model~\cite{moxnes2014using}).

%%%%%%%%%%%%%%%%%%%%%%%%%%%%%%%%%%%%%%%%%%%%%%%%%%%%%%%%%%%%%%%%%%%%%%
\subsection{Estimating the Braking Parameters}
\label{sec:braking}

The specification of the braking force in Eq.~\eqref{eq:Fb-threshold} is
incomplete without estimates for the braking coefficient $\gamma$ and
threshold $\acmin$.  The coefficient $\gamma$ can be determined using
experimental data reported in Sandbakk et
al.~\cite{sandbakk2014velocity}, who studied elite female skiers
applying both skid and step turn techniques while navigating curved
downhill sections of a ski course. These authors measured the
``trajectory'' ($D$, or distance travelled), average skier speed
($\overline{v}$), and the change in specific mechanical energy
($\Delta e$, or energy per unit mass) between the start/end of a turn.
Their data are summarized in the first four rows of
Table~\ref{tab:gamma} for four different combinations of turn geometry
($R$, $D$) and skid/step technique.

\begin{table}[bt]
  \centering
  \caption{Measured data from~\cite{sandbakk2014velocity} on elite
    female skiers (with average mass $m=60$~kg) performing several
    different skid/step turns. The first 4~rows list the original data
    while the last row contains our estimate of the braking coefficient
    $\gamma$ based on Eq.~\eqref{eq:gammaApprox}. Skiers employed the
    skid technique in the two tighter turns ($R=9,12$) and the step
    technique in wider turns ($R=12,15$).}
  {\renewcommand{\arraystretch}{1.2}
    \begin{tabular}{cc|cc|cc} 
      % \backslashbox[2cm]{\raisebox{0.1cm}{\mbox{~~}$\gamma$}}{$R=\frac{1}{\kappa}$} & 9 & 12 & 12 & 15  \\ [0.5ex] 
      Param. & Units & \multicolumn{2}{c|}{Skid turns} 
      & \multicolumn{2}{c}{Step turns}\\\hline
      $R=\ds\frac{1}{\kappa}$ & m & 9 & 12 & 12 & 15 \\
      $D$            & m & 16.5 & 21 & 21 & 24 \\
      $\overline{v}$ & m/s & 6.32 & 6.61 & 7.63 & 7.99 \\
      % $v_{in}$     & m/s & 7.58 & 7.41 & 7.51 & 7.62 \\
      $\Delta e$     & J/kg & 26.2 & 25.4 & 14.9 & 17.2 \\
      $\Delta E=m\Delta e$ & J & 1573 & 1525 & 892 & 1033 \\\hline
      $\gamma$       & -- & 0.3582 & 0.3324 & 0.1460 & 0.1686 \\[-0.5cm]
      && \multicolumn{2}{c|}{$\underbrace{\text{\hspace*{1cm}}}_{}$} 
      & \multicolumn{2}{c}{$\underbrace{\text{\hspace*{1cm}}}_{}$}\\[-0.4cm]
      && \multicolumn{2}{c|}{{\footnotesize$\gamskid\approx 0.3453$}} 
      & \multicolumn{2}{c}{{\footnotesize$\gamstep \approx 0.1573$}}
    \end{tabular}}
  \label{tab:gamma}
\end{table}

Assuming that the braking force is constant throughout each turn, the
mechanical energy (or work) can be estimated using the 
physical principle that ``work equals force times distance'':
\begin{gather}
  \underbrace{~\Delta E {\color{white}\big|}}_{\text{work}} 
  \;=\; 
  \Delta e \cdot  m
  \;=\; 
  \underbrace{\big( m \gamma \kappa \overline{v}^2 \big)}_{\text{force}} 
  \; \cdot \underbrace{~D{\color{white}\big|}}_{\text{distance}}
  \qquad \Longrightarrow \qquad 
  \gamma = \frac{\Delta e}{\kappa \overline{v}^2 D} 
  = \frac{R \,\Delta e}{\overline{v}^2 D} \raisebox{0.1cm}{.}
  \label{eq:gammaApprox}
\end{gather}
Substituting the average skier mass $m=60$~kg and other parameters from
Table~\ref{tab:gamma} into Eq.~\eqref{eq:gammaApprox} yields $\gamma$
values reported in the final row of the table.  Turning technique
clearly has a significant effect on the braking coefficient, with
$\gamma$ for the step turn equal to roughly half that for the skid
turn. On the other hand, $\gamma$ depends only very weakly on the turn
radius $R$ and length $D$, so we choose the average of the two braking
coefficient values $\gamskid=0.3453$ and $\gamstep = 0.1573$ for the
skid- and step-turn techniques, respectively.

The other parameter in our braking model is the threshold $\acmin$,
which cannot be determined from~\cite{sandbakk2014velocity} since they
did not consider the different situations when braking turn techniques
could be applied. However, the Homologation Manual provides a list of
maximum limits for centripetal
acceleration~\cite[p.~26]{FISHomologation-2021} that are permitted
within the various Nordic ski race formats.  The smallest two limits
allowed in all races are $a_c=2$ and 5~m/s${}^2$, which suggests two
possible values for $\acmin$ that will be investigated in the next
section.

%%%%%%%%%%%%%%%%%%%%%%%%%%%%%%%%%%%%%%%%%%%%%%%%%%%%%%%%%%%%%%%%%%%%%%
% --------------------------------------------------------------------
%%%%%%%%%%%%%%%%%%%%%%%%%%%%%%%%%%%%%%%%%%%%%%%%%%%%%%%%%%%%%%%%%%%%%% 
\section{Example 4: 3D Simulations of the Ole Course}
\label{sec:3Dsims}

We are now ready to simulate a skier moving along an actual 3D ski
course. We return to Example~3 and the 4.2~km Ole course from Toblach,
and solve the ODE system \eqref{eq:ode3a}--\eqref{eq:ode3b} by making
use of all Hermite spline computed by {\tt setup3d} (not just those for
$Z$, $S$ and $\Theta$). Six simulations are performed taking values of
the braking coefficient $\gamma=0$ (no braking), $0.3453$ (skid
  turn), and $0.1573$ (step turn), each being applied with two
threshold values $\acmin=2, 5$. Otherwise we use the same MSH baseline
parameters from Table~\ref{tab:params} that were used in 2D simulations.

The corresponding completion times are summarized in the first two rows
of Table~\ref{tab:3Dresults} (labelled ``Ole'') where either the skid or
step turn increases the finish time by roughly 10--30~s, which is a
significant difference in a race scenario.  The slowest time results
from applying the skid turn with threshold $\acmin=2$ (requiring an
extra 29.6~s), whereas increasing the threshold to $\acmin=5$
(reducing the frequency of braking) allows the skier to finish slightly
faster.  Note that the total time required without braking matches the
2D result of 819.9~s reported in Section~\ref{sec:2Dsims-Ole}, because
the 2D and 3D ODE systems are identical when $\gamma=0$.
For comparison purposes, we repeat these same six simulations using GPS
data from another 3.9~km FIS-rated course from Toblach called
``Stephanie,'' which features more tight curves than the Ole course. The
corresponding times are given in the last two rows of
Table~\ref{tab:3Dresults}, which exhibit time differences from the ``no
brake'' case that are more than double that from the Ole course. This
result is not surprising since the Stephanie course has more tight
curves on which braking forces are applied.

\begin{table}[htbp]
  \centering
  \caption{Simulated race times for the Ole and Stephanie courses in
    Toblach.  The results are presented for three choices of braking
    technique (no braking, skid turn, step turn) and two braking
    thresholds ($\acmin=2,5$). The figures in parentheses indicate the
    corresponding changes relative to the time without braking.}
  \renewcommand{\arraystretch}{1.2}
  \begin{tabular}{c|c|c|r@{~\;}c|r@{~\;}c} 
    Course  & $\acmin$ &  No brake 
    & \multicolumn{2}{c|}{Skid turn}
    & \multicolumn{2}{c}{Step turn} \\\hline
    \multirow{2}{*}{Ole (4.2~km)}
            & 2 & \multirow{2}{*}{819.9}
    & 849.5 & ($+$29.6) 
            & 837.5 & ($+$17.6) \\
            & 5 & & 837.6 & ($+$17.7)
            & 830.3 & ($+$10.4) \\\hline
    \multirow{2}{*}{Stephanie (3.9~km)}
            & 2 & \multirow{2}{*}{801.6}
    & 864.5 & ($+$62.9) 
            & 840.8 & ($+$39.2) \\
            & 5 & & 852.4 & ($+$50.8)
            & 834.3 & ($+$32.7)\\\hline
  \end{tabular}
  \label{tab:3Dresults}
\end{table}

\begin{figure}[tb]
  \centering
  % \begin{tikzpicture}
  %   \sffamily\Large\bfseries 
  %   \node at (0,0) {\includegraphics[width=0.6\textwidth]{fig9}};
  %   \draw (-3.1,2.4) node[color=brown] {A};
  %   \draw (2.35,2.4) node[color=brown] {B};
  %   \draw (3.3,2.4)  node[color=brown] {C};
  % \end{tikzpicture}
  \includegraphics[width=0.6\textwidth]{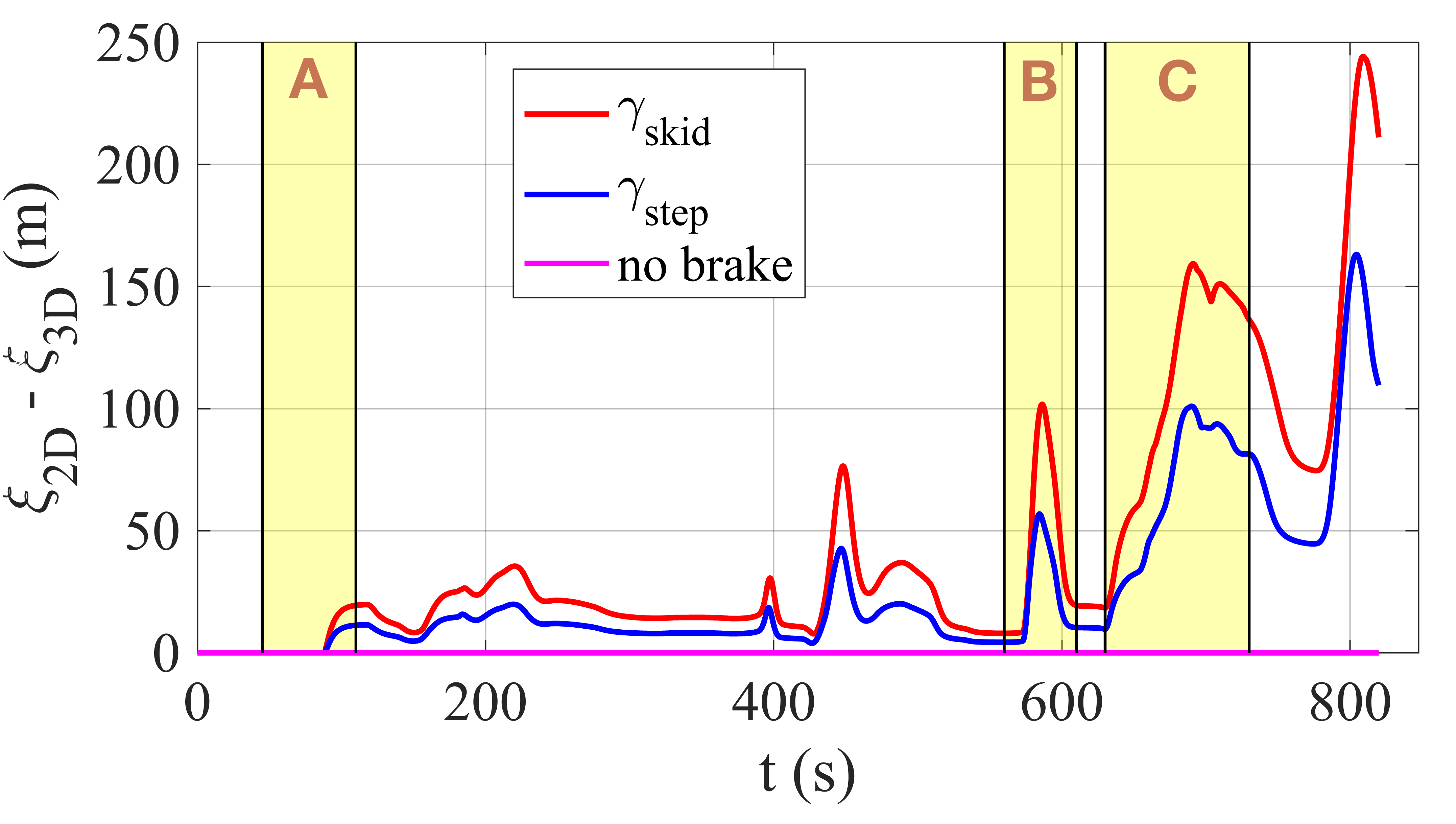}
  \caption{The difference in projected arc length $\xi$ between the 2D
    and 3D simulations on the Ole course with three choices of braking
    parameter ($\gamma=0, \gamstep, \gamskid$).  The braking threshold
    is set to $\acmin=2$ and all other parameters are taken from the MSH
    baseline test.}
  \label{fig:xi3Dvs2D}
\end{figure}

Returning to the Ole results, Fig.~\ref{fig:xi3Dvs2D} provides a more
detailed look at the effects of braking by plotting the difference in
projected distance, $\xi_{2D}-\xi_{3D}$, between the case with no
braking ($\xi_{2D}$) and that with skid/step turns ($\xi_{3D}$).  The
tightest downhill curves that occur in the final third of the course can
be identified with the largest discrepancies between the three
results. At the end of the course, the skier using skid turns finishes
roughly 240~m behind the idealized athlete who applies no braking,
whereas a skier using the more efficient step turns lags by only 120~m.
This emphasizes how a simple adjustment in braking technique on only a
few difficult turns could allow one skier to make significant gains over
another. As mentioned earlier in Section~\ref{sec:3Dbraking} (in
reference to Bj{\o}rgen's 2010 gold medal performance), even very small
differences can be of critical importance in a race scenario where
competitors are skiing neck-and-neck and must jockey for any possible
advantage.

Next, we focus on a single simulation with parameters
$\gamma=\gamma_{skid}$ and $\acmin=2$. The resulting braking force is
plotted on top of the elevation profile in Fig.~\ref{fig:ole-sim3D}a,
which underscores how the tightest curves are concentrated within the
second half of the course. Plots of the time variation of four solution
variables ($F_b$, $v$, $\kappa$ and $\inclination$) are also given in
Fig.~\ref{fig:ole-sim3D}b, which illustrate that a combination of large
speed and curvature is required to generate a significant braking
force. In the following discussion, we focus especially on the three
shaded regions of the course in Fig.~\ref{fig:xi3Dvs2D} that are
labelled A,B,C, and likewise highlighted in Fig.~\ref{fig:ole-sim3D}b:
\begin{figure}[!ht]
  \centering
  (a) Braking force along the 3D course \\
  \includegraphics[width=0.53\textwidth]{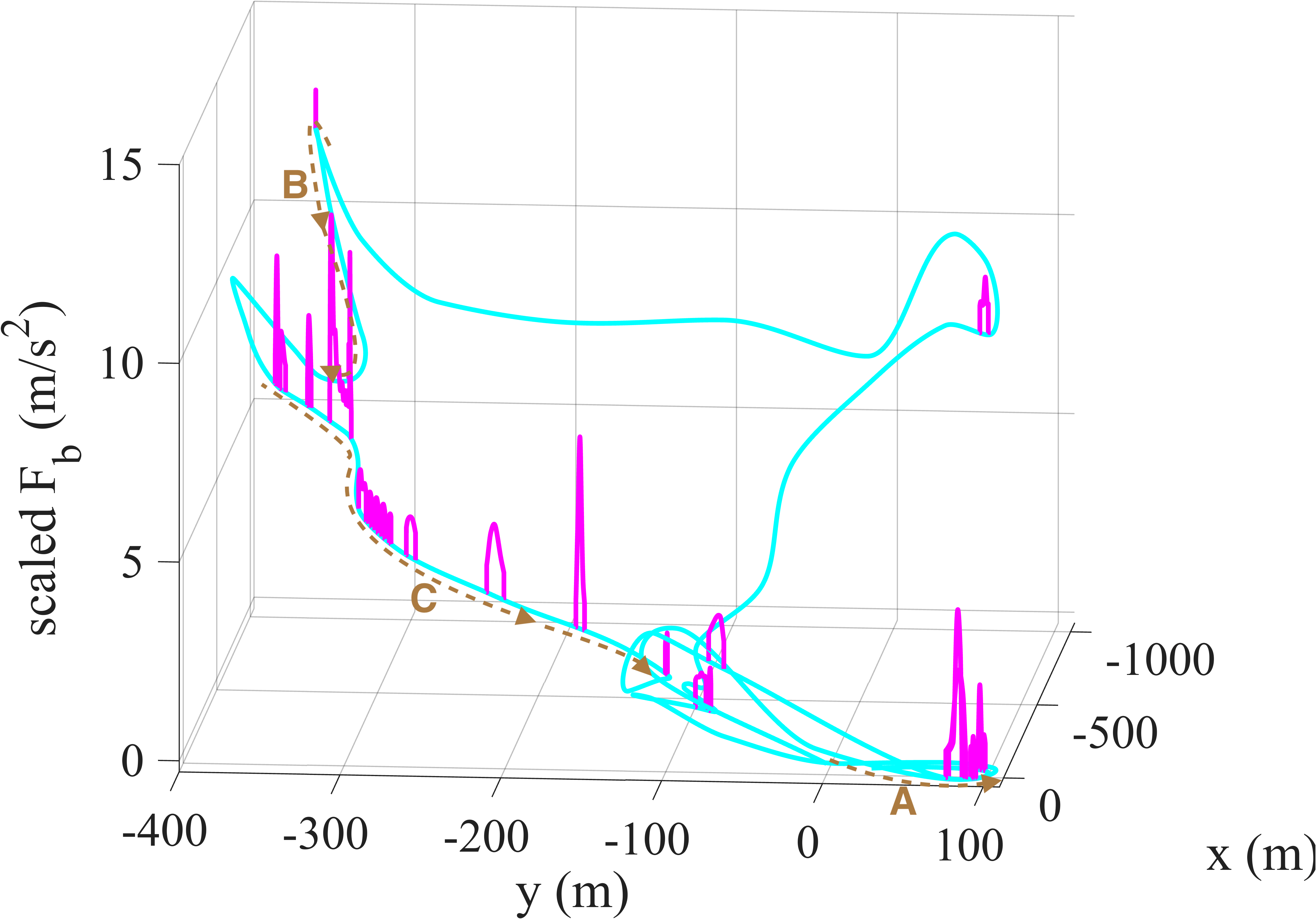}\\[0.2cm]
  \begin{tabular}{cc}
    (b) Solution components versus time
    & (c) Zoomed curvature in Region B\\
    %\begin{tikzpicture}
    %  \bfseries\sffamily
    %  \node at (0,0) {\includegraphics[width=0.47\textwidth]{fig10b_}};
    %  \draw (-2.4,2.9) node[color=brown] {A};
    %  \draw (1.9,2.9)  node[color=brown] {B};
    %  \draw (2.5,2.9)  node[color=brown] {C};
    %  \draw[color=black,thin,dashed] (1.6,-1.1)--(1.6,0.2)--(2.2,0.2)--(2.2,-1.1)--(1.6,-1.1);
    %\end{tikzpicture}
    \includegraphics[width=0.45\textwidth]{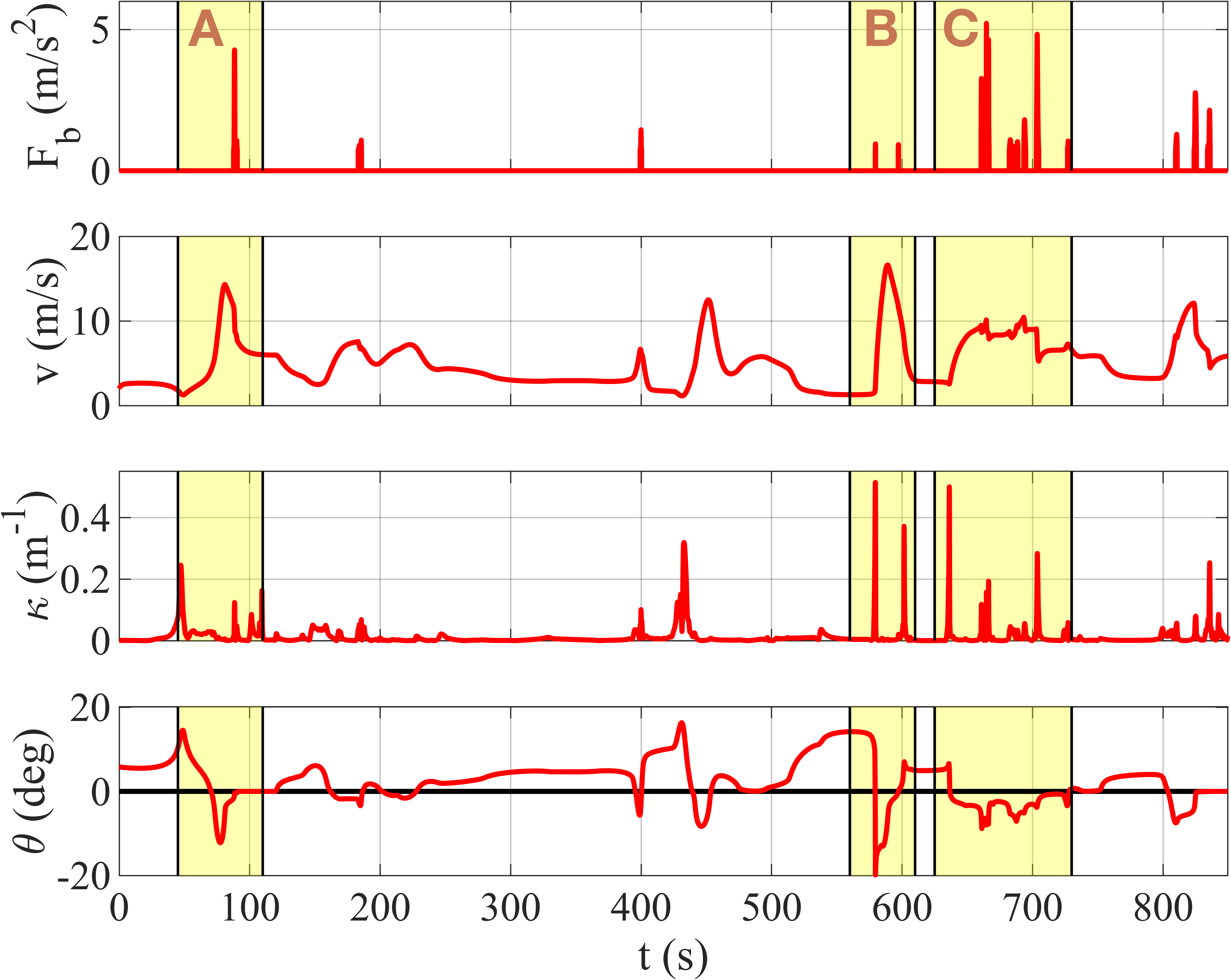}
    % & \raisebox{0.5cm}{\includegraphics[width=0.45\textwidth]{kappaZoomIn}}
      & \raisebox{0.5cm}{%
        \begin{tikzpicture}
          \node at (0,0) {\includegraphics[width=0.43\textwidth]{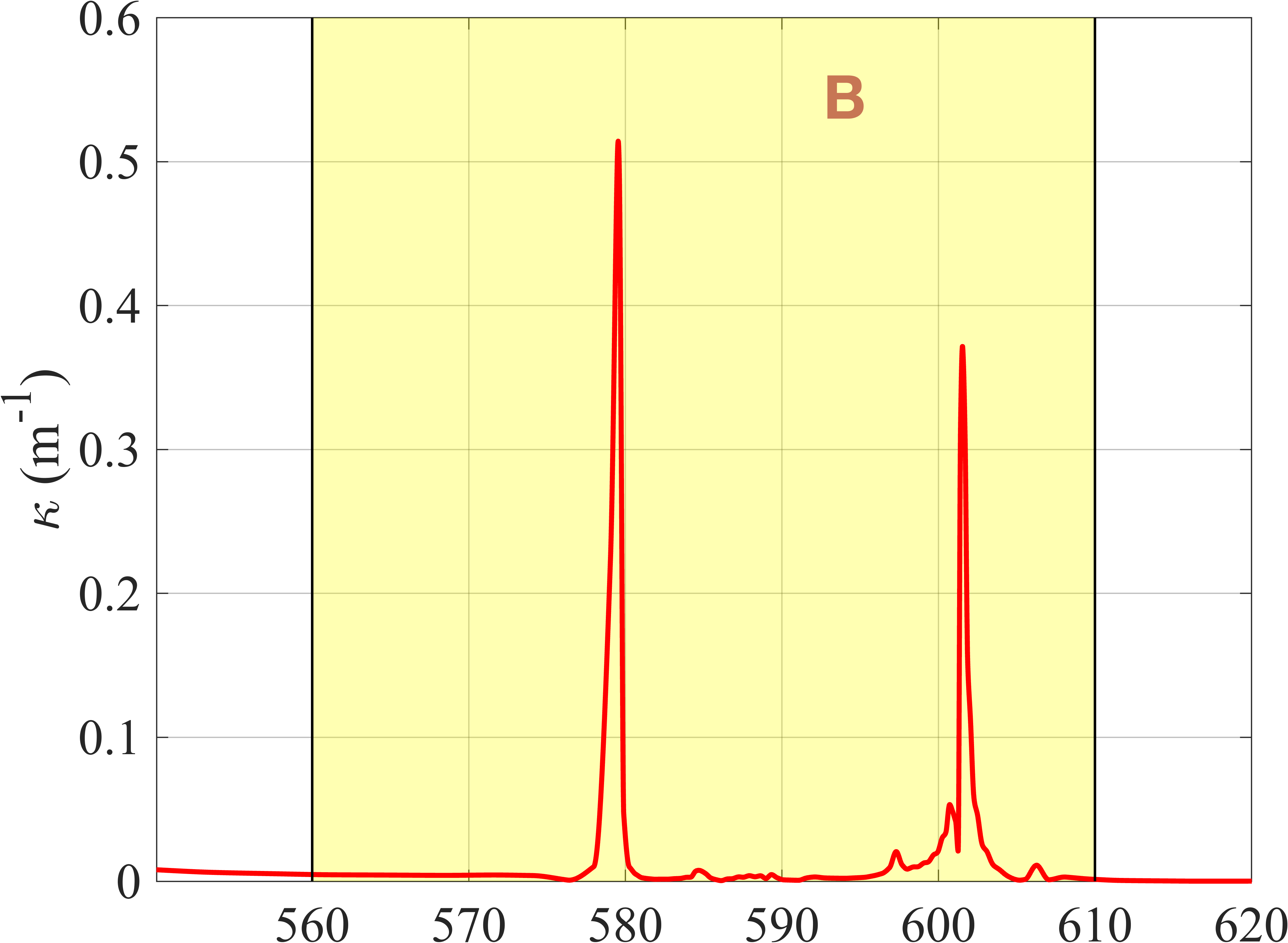}};
          \draw[very thick,color=brown,rotate around={0:(1.45,-2.2)}] (1.45,-2.2) ellipse (0.2 and 0.3);
          \draw (-0.2,2.1) node[color=brown!80!black] {$t\approx 580$};
          \draw (1.1,-1.75) node[color=brown!80!black]{$t\approx 597$};
          \draw (1.8,0.9) node[color=brown!80!black]  {$t\approx 602$};
        \end{tikzpicture}}
  \end{tabular}
  \caption{(a, Top) The braking force is displayed along the scaled
    Ole course profile in a 3D view of the solution with
    $\gamma=\gamskid$ and $\acmin=$2. The skier brakes much more often
    during the second half, where the course is mostly downhill.  The
      sections marked A,B,C indicate tightly curved downhill sections where
      braking forces are applied.
    (b, Bottom Left) Plots of $F_b$, $v$, $\kappa$ and $\inclination$
    versus time demonstrate how the braking force arises from a
    combination of skier speed and track curvature. (c, Bottom Right)
    Zoomed-in view of the curvature plot in region B.}
  \label{fig:ole-sim3D}
\end{figure}
\begin{itemize}
\item \emph{Region A:} This starts with a tight curve on an uphill climb
  at $t\approx 50$~s, followed by a steep downhill that also ends on a
  highly curved section. Here, the skier only brakes near the bottom of
  the hill because they are not travelling fast enough on the first
  short downhill section to exceed the braking threshold.

\item \emph{Region B:} This stretch contains the steepest downhill
  section on the course which also has two moderately tight curves near
  the hill crest and base, corresponding to the two prominent spikes in
  $\kappa$ near $t\approx 580$ and $602$~s. The first curve is
  encountered just over the hill crest when the skier is only beginning
  to speed up but their centripetal acceleration is still sufficient to
  exceed the braking threshold and generate a moderate braking force
  with $F_b\approx 1$. In contrast, the similar curvature near the base
  of the hill occurs once the course has flattened and the skier has
  slowed down enough that no braking is triggered. Nevertheless, there
  is a time in between (near $t\approx 597$~s) when the skier encounters
  only a modest curve on the steep downhill (circled in
  Fig.~\ref{fig:ole-sim3D}c), yet is moving sufficiently fast to induce
  a similarly-sized spike in $F_b$.  This example illustrates how
  braking forces can be generated by a very fast moving skier on a
  moderate curve, as well as on very tight curves at moderate speeds.
  
% \mymod{A similarly-sized spike in $F_b$ at $602$~s occurs on the
% second tight curve at the base where the skier has already slowed down
% significantly. On the remainder of the hill where the speed reaches a
% peak, the course is relatively straight and curvature is not large
% enough to induce any measurable braking force.  This example
% demonstrates how a skier travelling at moderate velocity may still
% generate a small braking force on especially tight curves.}
  
\item \emph{Region C:} This is a curvy and mostly downhill portion of
  the course, which generates the biggest collection of spikes in $F_b$
  where the braking threshold is exceeded. As a result, we also see a
  relatively large discrepancy in the distance in
  Fig.~\ref{fig:xi3Dvs2D} in this section.
\end{itemize}
An MPEG video of the 3D course is included in the Supplementary
Materials.

By adding curvature and a simple braking force into our model, we have
shown how to quantify the effects of applying various braking techniques
and that braking can have a significant impact on the finishing
time. Additionally on more convoluted courses like Stephanie, this
impact could be even greater.  This simple addition to the model
provides opportunities for collaboration with coaches, athletes, and
sports scientists to test our model predictions against real athlete
data, and to investigate different race strategies with a view to
optimizing race results.

%%%%%%%%%%%%%%%%%%%%%%%%%%%%%%%%%%%%%%%%%%%%%%%%%%%%%%%%%%%%%%%%%%%%%%
%\subsection{Exercises for the 3D Model}\label{sec:exercises3D}

%\begin{exercise}
%  \label{ex:gamma}
%  Show that $\gamma$, as introduced in Section~\ref{sec:3Dmodel} is
%  indeed a dimensionless quantity. Show that $\gamma$ as a
%  dimensionless quantity can be approximated by
%  Eq.~\eqref{eq:gammaApprox}. 
%\end{exercise}

%\begin{exercise}
%  \label{ex:trapezoidal}
%  Let $F(\alpha_k) = \sqrt{\hat{x}^2 + \hat{y}^2}$, where $\hat{x}$ and
%  $\hat{y}$ are finite difference approximations of the first
%  derivatives $x_\alpha$ and $y_\alpha$. Show that the trapezoidal rule
%  for approximating  
%  \begin{gather*}
%    \xi(\alpha_k) = \int_0^{\alpha_k} \sqrt{x_{\alpha}^2 + y_{\alpha}^2} d \alpha
%    \end{gather*}
%    reduces to $\sum_{j=1}^k \frac{1}{2} \big[ F(\alpha_j) +
%    F(\alpha_{j-1}) \big]$ when we assume constant spacing
%    $\Delta\alpha_j = \Delta\alpha$ for all $j$.
%\end{exercise}

\begin{exercise}
  \label{ex:beginner} 
  The emphasis so far has been overwhelmingly on expert skiers, so let's
  consider recreational or beginner Nordic skiers, by assuming these
  changes to the MSH baseline parameters:
  \begin{itemize}
  \item Beginners ski at much lower power levels, with $\Pmax$ being
    less than half the baseline value.
  \item They always remain upright, never switching to a tuck position
    on downhills.
  \item They use the ``snowplow'' technique on all downhills regardless
    of speed or curvature, employing the much larger braking parameter
    $\gamma=10\gamstep$.
  \end{itemize}
  For the Ole course, adjust the power parameter to obtain a
  finish time that is twice the baseline result.
\end{exercise} 

\begin{exercise}
  \label{ex:albertSprint3D}
  Download GPS data for the 1.4~km ``Albert Sprint'' course from the
  Dolomiti NordicSki website~\cite{dolomiti-gps-2023} and simulate a
  skier in 3D with curvature and braking. Compare your results with
  those from the corresponding 2D simulation in
  Exercise~\ref{ex:albertSprint}.
\end{exercise}

\begin{exercise}
  \label{ex:starcourse}
  Consider an artificial 3D course that is built using the function
  $r(\theta) = 100(1+\frac{1}{2}\cos n\theta)$, which describes a curve
  in the $x,y$-plane in polar coordinates (for $n\geqslant 1$ an
  integer). To introduce changes in elevation, this curve is then
  rotated an angle of $10^\circ$ about the $y$-axis. This yields a
  smooth ``star-shaped'' path having $n$ rounded points or lobes that
  become more highly curved as $n$ increases, and with gradients varying
  between $\pm 10^\circ$. The Supplementary Information includes a
  \Matlab\ function called {\tt starcourse.m} that generates $(x,y,z)$
  coordinates using:\\ 
  \mbox{\hspace*{0.5cm}}
  {\tt [x, y, z] = starcourse(3, 100)} \\
  where the first argument is the number of lobes (here, $n=3$) and the
  second argument indicates that 100 data points should be used to
  define the course. Modify the code {\tt setup3d.m} to use this command
  to define the course rather than reading a GPS data file, and then run
  the code for a selection of $n$ values between 1 and 20. The code
  will generate some anomalous or erroneous results for different
  values of $n$. Discuss the results, explain what is going wrong when
  errors arise, and then try to correct the code.
\end{exercise}

% \begin{exercise}
%   Other similar models for Nordic skiing include the effect of
%   vertical curvature in the $\xi,z$-plane~\cite{carlsson-etal-2011,
%   moxnes2014using, sundstrom2013numerical}, which affects the forces
%   due to both gravity and snow friction. Show that this is negligible
%   in relation to sideways curvature on downhill turns.
% \end{exercise}

%%%%%%%%%%%%%%%%%%%%%%%%%%%%%%%%%%%%%%%%%%%%%%%%%%%%%%%%%%%%%%%%%%%%%%
% ----------------------------- ---------------------------------------
%%%%%%%%%%%%%%%%%%%%%%%%%%%%%%%%%%%%%%%%%%%%%%%%%%%%%%%%%%%%%%%%%%%%%% 
\section{Discussion}
\label{sec:discussion}

We have developed an ODE-based model that captures the essential
dynamics of Nordic skiing, and provides a stimulating example of the
mathematical modelling process that can be easily incorporated into
undergraduate mathematics classes.  This model is an excellent
illustration of the inherent complexity of \emph{real problems}, which
invariably require a multi-pronged modelling strategy that combines
concepts and techniques from several different areas including:
\begin{itemize}
\item \emph{Data analysis and cleaning:} 
  % ALSO: data scrubbing
  to investigate and correct for missing data points, highly non-uniform
  point spacing, and other measurement errors that are unavoidable with
  real data sources.
        
\item \emph{Function approximation:} using piecewise spline
  interpolation to construct smooth representations for sparse data.
    
\item \emph{Ordinary differential equations:} that are derived by
  combining physical intuition with basic laws of Newtonian mechanics.
    
\item \emph{Numerical algorithms:} to approximate the governing
  equations and validate the model predictions.
    
\item \emph{Scientific visualization:} although this was not explicitly
  addressed, we deliberately employed a varied selection of graphical
  representations for 2D and 3D data, with the intent to accentuate any
  differences and to provide extra insight into the results.
\end{itemize}

Besides the educational value of this skiing model in a classroom
setting, we have also attempted to demonstrate how even fairly
elementary mathematical techniques can stimulate novel advances in
sports science. One example is our use of Hermite splines to approximate
a ski course, which provides an attractive balance between providing
smoothness and avoiding spurious oscillations. There are also very few
3D models of Nordic skiing, and ours is the first to incorporate the
inherently 3D effect of braking in tight downhill curves, which are
important strategic sections of competitive races. Examples like these
can be exploited to motivate students that the mathematics they are
learning has practical value in solving real problems, but can also lead
to advances in fundamental knowledge.

Another significant contribution of this paper is our open-source
\Matlab\ code, which is flexible enough to simulate skiers on both 2D
and 3D course geometries, and handles multiple input formats for the
course data. Because \Matlab\ is so widely used in university settings,
access to such a code should allow instructors to easily incorporate
this material into their courses, while also providing students with a
realistic and stimulating problem to experiment on and gain insight
from. Besides its educational uses, we also anticipate that this skier
simulation code will be of benefit to sports science researchers, who
are free to experiment with and generalize the underlying algorithms for
their own purposes.

In closing, we should emphasize that this work is still only a
preliminary step towards developing a truly realistic model for a very
complex and multi-faceted sport, and there are a host of opportunities
for further work. One specific area is the study of athlete fatigue in
medium- to long- distance events, and how to incorporate experimental
observations on positive pacing into the power model beyond the na\"ive
approach suggested in Exercise~\ref{ex:fatigue}.  Another natural
extension is to generalize our simplistic approach for handling braking
turns to incorporate ``look-ahead,'' where a skier anticipates arriving
at a tight curve by initiating their braking technique well in advance.
There are many other fascinating aspects of Nordic skiing that we have
so far ignored, such as the metabolic processes behind muscle power
generation, or the variability in course conditions due to
weather-induced snow conditions, or track grooming, or athlete race
strategy, to name just a few. In terms of strategic decision-making,
mathematics is especially well-suited to drive advances in understanding
how specific choices of technique or training regimen could optimize
race performance under a variety of parameter regimes. Finally, existing
models of athlete dynamics in other competitive sports could provide
interesting avenues to explore in the context of Nordic skiing. For
example, optimization methods developed to maximize performance in a
running race~\cite{aftalion2014optimization, keller1974optimal} could be
extended to skier dynamics along similar {lines} to what was already
done in~\cite{Driessel2004dynamics}. Furthermore, in the sport of road
cycling athletes propel themselves at much higher speeds where
aerodynamic forces dominate~\cite{nee-herterich-2022}, which would
provide an opportunity to draw interesting comparisons with skiing.

\section*{Acknowledgements}

We thank Kayden Sim (GIS Specialist, Forsite Consultants) and Jake
Weaver (Head Coach, Hollyburn Cross-Country Ski Club, North Vancouver,
BC). We also appreciate the advice of FIS Technical Delegates John
Aalberg (Sooke, BC) and Al Maddox (Atikokan, ON) who assisted us with
details regarding course homologation.
Finally, we are very grateful to the two anonymous reviewers for their
extensive and insightful comments that allowed us to make significant
improvements to our final paper, and provided the inspiration for two
additional exercises.

%% 
%% The next two sections were adapted from a recent SIREV article 
%% on "Combinatorial and Hodge Laplacians"
%%
%\section*{Code Availability}

\section*{Supplementary Materials}
\begin{itemize}
\item[S1.] Exercise solutions (suppl1Solutions.pdf, 11.9~Mb)
\item[S2.] Summary of \Matlab\ codes and data files (suppl2CodesAndData.pdf, 131~Kb)
\item[S3.] \Matlab\ codes (suppl3Matlab.zip, 15~Kb)
\item[S4.] Data files (suppl4DataFiles.zip, 13~Kb)
\item[S5.] Movie of 2D MSH course simulation (suppl5Movie2dMSH.mp4, 2.1~Mb)
\item[S6.] Movie of 3D Ole course simulation (suppl6Movie3dOle.mp4, 2.7~Mb)
\item[S7.] Movie of 3D simulation for ``star-shaped'' course in
  Exercise~\ref{ex:starcourse} (suppl7Movie3dStar7.mp4, 2.0~Mb)
\end{itemize}

%%%%%%%%%%%%%%%%%%%%%%%%%%%%%%%%%%%%%%%%%%%%%%%%%%%%%%%%%%%%%%%%%%%%%%
% --------------------------------------------------------------------
%%%%%%%%%%%%%%%%%%%%%%%%%%%%%%%%%%%%%%%%%%%%%%%%%%%%%%%%%%%%%%%%%%%%%% 

%\bibliographystyle{abbrv}
%\bibliography{bibli}
%\end{document}

\end{document}